\RequirePackage{silence}
\documentclass[aps,pra,twocolumn,showpacs,amsmath,amssymb,preprintnumbers,superscriptaddress,10pt,longbibliography]{revtex4-2}

\usepackage{graphicx}
\graphicspath{{figures/}}
\usepackage[binary]{SIunits}
\usepackage[bookmarks=false]{hyperref}
\usepackage{color}
\usepackage[capitalise]{cleveref}
\crefname{section}{Sec.}{Secs.}
\Crefname{section}{Section}{Sections}
\usepackage{xurl}

\definecolor{pink}{RGB}{255,0,255}

\definecolor{ginger}{RGB}{255,150,0}

\definecolor{pine}{RGB}{1,121,111}

\definecolor{RED}{RGB}{255,0,0}

\begin{document}

\title{Energy--time attack on detectors in quantum key distribution}

\author{Konstantin~Zaitsev}
\email{zaitsev20k@gmail.com}
\affiliation{Vigo Quantum Communication Center, University of Vigo, Vigo E-36310, Spain}
\affiliation{atlanTTic Research Center, University of Vigo, Vigo E-36310, Spain}
\affiliation{Russian Quantum Center, Skolkovo, Moscow 121205, Russia}
\affiliation{NTI Center for Quantum Communications, National University of Science and Technology MISiS, Moscow 119049, Russia}

\author{Vladimir~Bizin}
\affiliation{Russian Quantum Center, Skolkovo, Moscow 121205, Russia}
\affiliation{Moscow Institute of Physics and Technology, Dolgoprudny, Moscow region 141700, Russia}

\author{Dmitriy~Kuzmin}
\affiliation{Moscow Technical University of Communications and Informatics, Moscow 111024, Russia}
\affiliation{QRate, Skolkovo, Moscow 143025, Russia}
\affiliation{NTI Center for Quantum Communications, National University of Science and Technology MISiS, Moscow 119049, Russia}

\author{Vadim~Makarov}
\affiliation{Russian Quantum Center, Skolkovo, Moscow 121205, Russia}
\affiliation{Vigo Quantum Communication Center, University of Vigo, Vigo E-36310, Spain}
\affiliation{NTI Center for Quantum Communications, National University of Science and Technology MISiS, Moscow 119049, Russia}

\date{May~12, 2026}

\begin{abstract}
Quantum key distribution is unbreakable in theory but may be hacked via imperfections in its hardware implementations. While many imperfections have been mitigated by countermeasures and advanced security proofs, several remain unsolved. One of these is a superlinear behaviour in single-photon detectors, when the click probability rises faster with the photon number of an incoming light pulse than expected from individual independent photon detections. Here we test an avalanche single-photon detector sinusoidally-gated at 312.5~MHz for superlinearity. Its click probability is moderately superlinear. However, we notice that the click timing depends strongly on the incoming pulse energy. The click occurs progressively earlier, shifting more than 2~ns as the energy rises over a wide 50-dB range. An attacker might use this energy--time effect to conditionally toggle the click between adjacent key bit slots, violating an implicit assumption in the security proofs and rendering them inapplicable. We propose two attacks that exploit this flaw.
\end{abstract}

\maketitle

\section{Introduction}
\label{sec:intro}

Security of data sharing including private communications, financial transactions, and medical information play an important role in modern life. With the second quantum revolution underway \cite{dowling2003}, the security question rises again. Widely used cryptography algorithms based on computation complexity of factorization \cite{rivest1978} may be hacked by quantum computers in near future \cite{shor1997}. To avoid the collapse of information security, quantum key distribution (QKD) technology is proposed \cite{bennett1984}. Its security is based on quantum mechanics and can not be broken, in theory. Unfortunately, the first QKD algorithms assumed that the optical hardware, on which they are implemented, is perfect. Further investigations have shown that this condition cannot be met at the current technology level. Multiple attacks on QKD exploiting imperfections of its hardware implementations have been proposed \cite{brassard2000,makarov2006,zhao2008,xu2010,lydersen2010a,gerhardt2011,sun2011,jain2011, weier2011,li2011a,jiang2012,jouguet2013,jain2014,sajeed2015,rau2015,sajeed2015a,makarov2016,yoshino2018,huang2018,huang2019,huang2023,ye2023,lu2023,baliuka2023}. 

One approach to fix this problem is to develop device-independent algorithms \cite{acin2006,briegel1998}. They have shown an impressive progress, particularly in measurement-device-independent QKD \cite{lo2012,tang2014}, source-independent QKD \cite{xiongfeng2007}, source-independent quantum random number generators \cite{zhu2016}, and quantum teleportation \cite{gottesman1999,bouwmeester1997}. The price for their higher security is a high complexity of the QKD setup and a dramatic decrease of the key rate. Another approach is a proper certification of imperfect single-photon sources and detectors with countermeasures against well-known threats. This allows the use of a cheaper prepare-and-measure high-rate QKD \cite{bennett1984} at the price of complicating its certification procedure. An example of the second approach can be found in \cite{makarov2024}, where a QKD system manufactured by QRate is analysed. However this approach is based on the knowledge of possible attacks, which is not guaranteed to be complete. Starting from year 2000, when the first attack was discussed in the literature \cite{brassard2000}, quantum hackers find a new attack once a year, on average \cite{brassard2000,makarov2006,zhao2008,xu2010,lydersen2010a,gerhardt2011,sun2011,jain2011,weier2011,li2011a,jiang2012,jouguet2013,jain2014,sajeed2015,rau2015,sajeed2015a,makarov2016,yoshino2018,huang2018,huang2019,huang2023,ye2023,lu2023,baliuka2023}. Notably, a superlinear behaviour has been observed in a wide range of single-photon detectors \cite{lydersen2010a,gerhardt2011,wiechers2011,lydersen2011b,lydersen2011c,chaiwongkhot2022}.

Here we report a loophole that threatens the security of prepare-and-measure QKD. We have found that a single-photon detector (SPD) attacked by a short bright pulse responds to a higher-energy pulse faster compared to a lower-energy pulse. We call this an energy--time effect. The total time shift observed exceeds $2~\nano\second$. This effect can be a tool for an eavesdropper Eve to manipulate the registration time of click events. This is accounted for in neither security proofs \cite{tsurumaru2008,xu2020}, nor in current certification standards \cite{iso23837-2023,marquardt2023}, nor in the recent analysis \cite{makarov2024}.

This work is part of an effort to design a secure singe-photon detector for QKD implementations. We expect that efficiency-mismatch attacks might be closed by implementing a four-state Bob (see \cite{makarov2024}), blinding attacks might be closed by a photocurrent measurement (see \cite{acheva2023}), and superlinearity can be characterised, as we also demonstrate in this paper. We find that the approach to characterising superlinearity discussed in \cite{lydersen2011b} needs an improvement for certification, which we implement. We compare our improved method to the original one and find it to be applicable to a much wider range of attack parameters under test. For the particular detectors tested in this paper, we observe a limited amount of superlinearity, which may or may not be sufficient for an after-gate attack \cite{wiechers2011}.

The paper is organised as follows. We describe the experimental setup and detector under test in \cref{sec:setup}. In \cref{sec:results}, we present the methodology of superlinearity measurement and its results, characterise the energy--time effect, and also notice a memory effect. In \cref{sec:attacks}, we propose possible attacks on QKD systems exploiting the energy--time effect. We discuss countermeasures in \cref{sec:countermeasures} and conclude in \cref{sec:conclusion}.

\section{Experimental setup}
\label{sec:setup}

\subsection{Detector under test}
\label{sec:DUT}

We test two samples (SPD1 and SPD2) of a detector designed by QRate. Its scheme and work principles are discussed in \cite{losev2022}. The detector uses an InGaAs/InP single-photon avalanche diode with a mixed passive-active quenching, which provides a dead time of $4.34~\micro\second$ (SPD1) and $4.36~\micro\second$ (SPD2).

The detector is sinusoidally-gated at a frequency of $F = 312.5~\mega\hertz$ (period of $3.2~\nano\second$). It is synchronised by an external $100$-$\mega\hertz$ clock input. The gating frequency of $312.5~\mega\hertz$ is formed internally by multiplying the $100~\mega\hertz$ clock by 25/8 inside the detector.

The photon detection efficiency (measured at 0.1 photons per attenuated laser pulse applied at $1~\kilo\hertz$ rate) is 9.8\% and 12.7\% for SPD1 and SPD2, while their dark count rate $D = 428$ and $532~\hertz$. To reduce the influence of the dark count rate on the measurement of photon detection efficiency, we use a coincidence logic detailed below.

One can see that the sample parameters do not perfectly match. If a pair of such SPDs is used in a QKD receiver, its security can be provided by amending the QKD protocol \cite{bochkov2019,trushechkin2022}. Alternatively, the four-state Bob cancels any mismatch between the two SPDs, including time- and wavelength-mismatch \cite{makarov2024}.

\subsection{Testbench and methodology}
\label{sec:testbench}

\begin{figure}
	\includegraphics{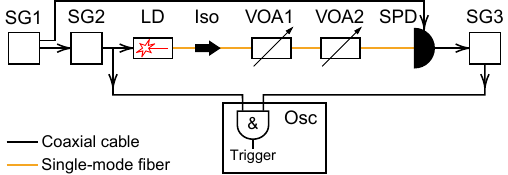}
	\caption{Experimental setup. SG, signal generator (SG1, Keysight 81180B; SG2 and SG3, Highland Technology P400); LD, laser diode (Gouch \& Housego AA1405-193200-100-PM900-FCA-00); Iso, isolator (Thorlabs IO-H-1550APC); Osc, oscilloscope (LeCroy WavePro 735Zi); VOA, variable optical attenuator (OZ~Optics DA100); SPD, single-photon detector under test (QRate).}
	\label{fig:setup}
\end{figure}

In order to investigate the detector's superlinearity and energy--time effect mentioned above, we use a scheme shown in \cref{fig:setup}. The first signal generator (SG1) serves as a clock sending $100$-$\mega\hertz$, $3$-$\volt$ peak-to-peak sine wave to the SPD and second signal generator (SG2). The latter sends two simultaneous pulses at $f = 1~\kilo\hertz$ repetition rate: one to drive a gain-switched $1551$-$\nano\meter$ laser diode LD, the other to oscilloscope Osc. The laser diode produces light pulses of $269$-$\pico\second$ full-width at half-maximum and is followed by an isolator Iso and two variable optical attenuators VOA1 and VOA2, which allow for sufficient optical attenuation up to $120~\deci\bel$. Attenuated optical pulses are sensed by the SPD and its output is fed into the third signal generator (SG3), which resends the signal into the oscilloscope. In the oscilloscope, signals from SG2 and SG3 are logically ANDed using the oscilloscope's software and the resulting signal is used as a trigger. This scheme selects click events caused by laser pulses in a $\theta = 13$-$\nano\second$ coincidence window, excluding most of dark clicks and afterpulses. The remaining effective dark count rate is under $10^{-2}~\hertz$. The timing of the laser pulses relative to the SPD's clock input is controlled by SG2, thus enabling the measurement to be conducted at any temporal point of the SPD's gate. The width of the coincidence window $\theta$ is adjusted by the width of pulses coming from SG2 and SG3 into Osc.

This setup is employed to register the SPD's response to the optical pulses. Two particular characteristics of the response are of interest: count rate and click time of the SPD relative to the moment the laser pulse impinges it. These characteristics are measured at 8 different points of the gate, $400~\pico\second$ apart. The first point is at the gate's maximum single-photon detection efficiency, measured with a highly attenuated optical pulse. At each point, the optical pulse energy is varied in a wide range of up to $102~\deci\bel$ with $3$-$\deci\bel$ step. The recorded count rate is used to characterise the SPD's superlinearity, while the click timing distribution characterises the energy--time effect.

\section{Experimental results}
\label{sec:results}

\subsection{Superlinearity measurement}
\label{sec:superlinearity}

\begin{figure*}
	\includegraphics{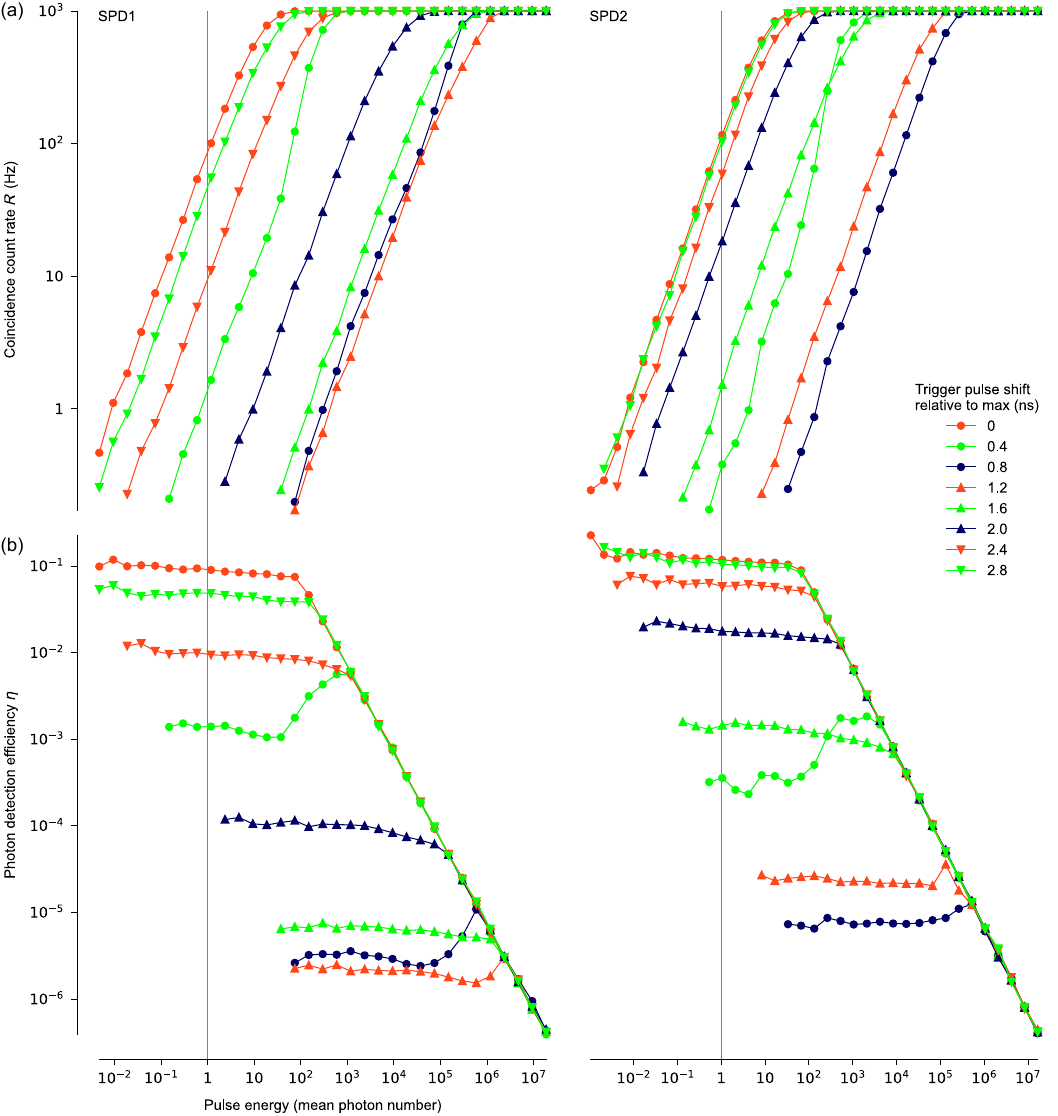}
	\caption{Superlinearity characterisation for SPD1 and SPD2. (a)~Coincidence count rate $R$. (b)~Photon detection efficiency $\eta$ calculated from \cref{eq:efficiency}. A straight line with negative slope at high energy is caused by saturation. Note that with true saturation, $R \ge f$ and $\eta(t, \mu)$ can not be evaluated. In our case, dark counts and afterpulses sometimes cause the detector's being in a deadtime during the coincidence window, resulting in $R < f$ at saturation. There, $R$ is virtually independent of $\mu$ and $\ln \eta(t, \mu)$ becomes an almost linear function of $\ln \mu$.}
	\label{fig:superlinearity-characterisation}
\end{figure*}

The concept of superlinearity measurement is suggested in \cite{lydersen2011b} for  gated threshold detectors. Gated detectors have a periodically repeating profile of single-photon detection efficiency $\eta(t)=\eta(T+t)$, where $T$ is gate period. Let's consider a perfect noiseless detector. Its detection probability of an $n$-photon pulse would be
\begin{equation}
	\label{eq:detector-linear-expectation}
	p_\text{det}(t,n)=1-\left[1-\eta(t)\right]^n
\end{equation}
if the photons are detected independently. In a coherent state, the number of photon per pulse $n$ follows Poisson distribution $p_n=\mu^ne^{-\mu}/n!,$ where $\mu$ is the mean photon number. Detection probability can thus be rewritten as 
\begin{equation}
	\label{eq:Poisson_case}
	p_\text{det}(t,\mu)=\sum_{n=0}^{\infty} \frac{\mu^ne^{-\mu}}{n!}p_\text{det}(t,n)=1-e^{-\mu\eta(t)}.
\end{equation}
Any detection probability higher than that is defined as superlinearity \cite{lydersen2011b}.

While this approach is clear, it has two significant drawbacks. First, it requires measuring the detector photon count rate at $\mu < 1$, which can be very low at trigger pulse shifts in-between the gates [see our experimental data in \cref{fig:superlinearity-characterisation}(a)]. This does not allow to determine $\eta(t)$ there, as the data primarily consists of residual dark counts. Second, it may miss superlinearity, as explained below.

To avoid these problems, we redefine $\eta(t, \mu)$ as a function of $\mu$. The coincidence count rate $R$ can be expressed as
\begin{equation}
	\label{eq:click-rate}
	R(t, \mu) = f (1 - e^{- \eta(t, \mu) \mu}) + f e^{- \eta(t, \mu) \mu} \left[1 - (1 - P_D)^{\theta F}\right],
\end{equation}
where $P_D = D/F$ is the dark count probability per gate. Here, the first term is a contribution from photon clicks and the second term is that from dark counts (the window is about $\theta F \approx 4$ gates long in our experiment). Solving for $\eta$, we get
\begin{equation}
	\label{eq:efficiency}
	\eta(t,\mu) = \frac{1}{\mu} \ln \frac{(1 - P_D)^{\theta F}}{1 - R(t, \mu) / f}.
\end{equation}
We now define that the detector is superlinear if
\begin{equation}
	\label{eq:new-superlinearity-def}
	\exists\ \mu_1, \mu_2, t : \mu_1 < \mu_2\ \text{and}\ \eta(t, \mu_1) < \eta(t, \mu_2).
\end{equation}

We believe that this definition is more general. Consider a case when $\mu$ increases and the detection efficiency first drops, then recovers. Such behaviour is observed, although faintly, in SPD1 at trigger pulse shifts of $0.4$, $0.8$, and $1.2~\nano\second$ [see \cref{fig:superlinearity-characterisation}(b)]. It is reasonable to call this behaviour superlinear, because it may be exploited in an after-gate attack \cite{lydersen2011b}. However, the original definition of superlinearity might fail to identify it as such.

The new definition allows to find points of superlinear behaviour of SPDs, but it does not allow to quantify the effect. We suggest to quantify the superlinearity by a partial derivative
\begin{equation}
	\label{eq:superlinearity-factor}
	S = \frac{\partial \ln \eta(t, \mu)}{\partial \ln \mu}.
\end{equation}
Its value describes by what factor $\eta$ changes, given that $\mu$ changes by some small factor. If $\eta(t, \mu)$ is continuously differentiable, the condition \labelcref{eq:new-superlinearity-def} is equivalent to
\begin{equation}
	\exists\ \mu, t : S > 0.
	\label{eq:new-superlinearity-def-alt}
\end{equation}
In the ideal SPD, $\eta$ does not depend on $\mu$, thus $S = 0$. A calculation of $S$ with a central difference formula yields a maximum value of 0.86 (0.90) for SPD1 (SPD2). This value can be interpreted as follows: if $\mu$ increases by a factor of $2$, then $\eta$ of SPD1 (SPD2) increases by a factor of about $2^{0.86}\approx 1.8$ ($2^{0.90}\approx 1.9$). 

\begin{figure*}
	\includegraphics{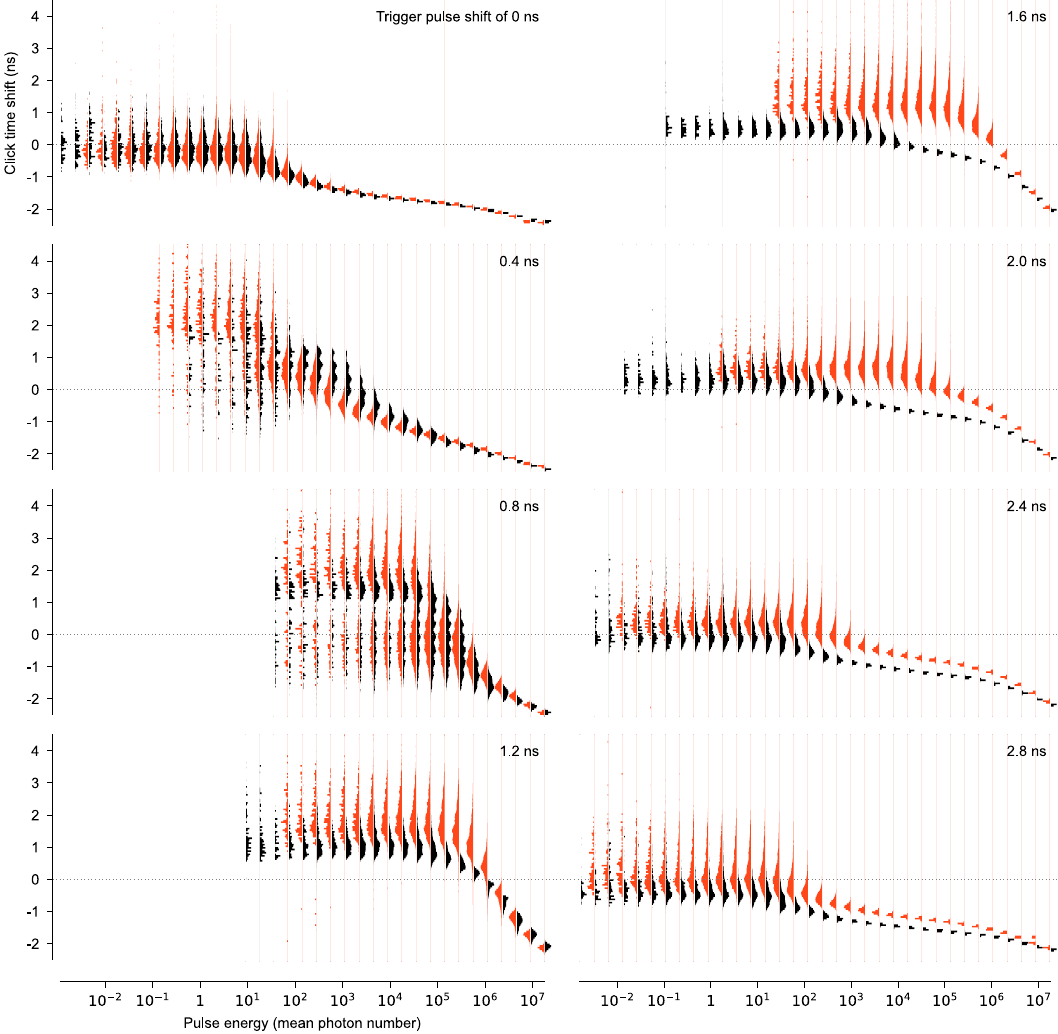}
	\caption{Click time shift distribution for each detector: red (gray) for SPD1 and black for SPD2. Each histogram is normalised individually.\\} 
	\label{fig:reaction-time}
\end{figure*}

The advantage of the new definition of superlinearity and its quantifying factor $S$ is their universality. However, their relation with the security of QKD requires further theoretical analysis. We presently only know how to calculate a quantum bit error rate (QBER) in an intercept-resend attack \cite{lydersen2011b}, which may not be the optimal attack:
\begin{equation}
	\label{eq:old_qber}
	\text{QBER} = \frac{2p_h - p_h^2}{2p_f + 2(2p_h - p_h^2)},
\end{equation}
where $p_f$ is the detection probability of a pulse of a certain energy and $p_h$ is that of half the energy, assuming they are the same for both detectors. The optimum pulse energy and trigger pulse shift for this attack does not necessarily coincide with those for the largest $S$. For our detectors, QBER of this attack $\geq 16.7\%$, which is too high for key generation and makes this attack fail.

In summary, we have measured a limited amount of superlinearity in this detector model. It remains unclear whether Eve would be able to exploit it for a working attack.

\begin{figure*}
	\includegraphics{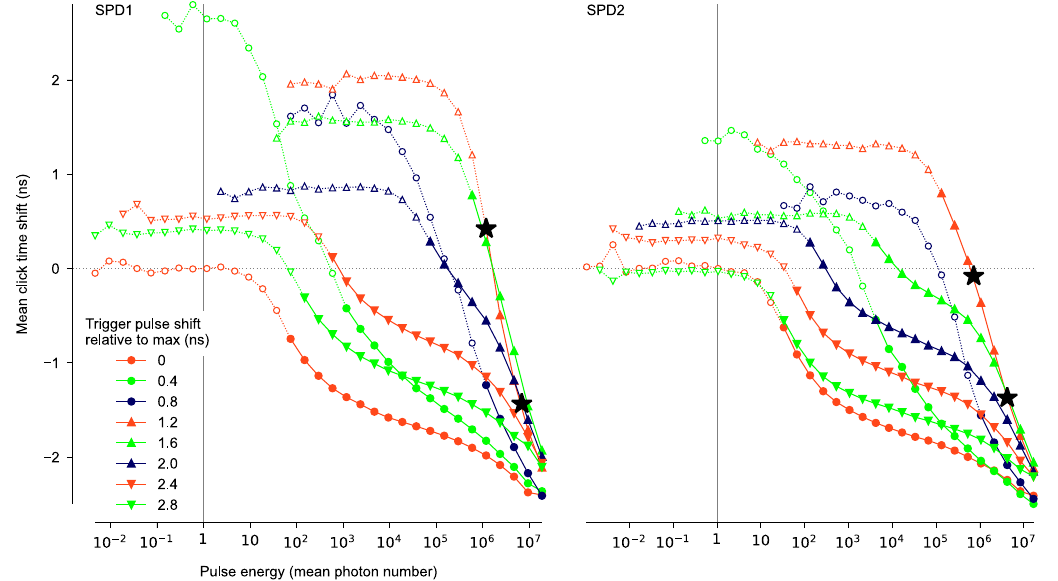}
	\caption{Mean value of click time shift distributions in \cref{fig:reaction-time}. Dotted lines and hollow symbols denote full-width at half-maximum (FWHM) of distribution exceeding $0.32~\nano\second$ (i.e.,\ one-tenth of the gate period), indicating qualitatively that it would be more difficult for Eve to control the click's bit slot. Conversely when the distribution is narrow, denoted by solid symbols and lines, she may enjoy near-deterministic control over the click timing. Black stars indicate conditions for each detector that maximise the difference in its click timing for an energy ratio of $7.7~\deci\bel$; they occur at $1.2$-$\nano\second$ trigger pulse shift for both detectors.}
	\label{fig:reaction-mean}
\end{figure*}

\subsection{Energy--time effect}
\label{sec:energy-time}

The delay between pulses generated by SG2 and SG3 is measured. We will refer to this delay as click time shift. The results obtained are presented in \cref{fig:reaction-time} as eight sets of histograms for different points of the gate. The click time shift of zero corresponds to the mean value of the histogram at $\mu = 1$ and trigger pulse shift of $0~\nano\second$, i.e., the timing of detector's normal single-photon response.

As can be seen, the distribution becomes much narrower and shifts to earlier time at higher energy. The mean click time shift for the eight points of the gate is plotted in \cref{fig:reaction-mean}. Each curve spans $2~\nano\second$ or more. The lack of saturation of the time shift at high energy hints that the limit of this effect has not been reached in our test.

Some interesting observations can be made looking at \cref{fig:reaction-time}. The first one is the presence of delayed detection events, which can be seen in distributions taken at $0.4$ and $0.8$-$\nano\second$ trigger pulse shifts. Sometimes the optical pulse sent after gate maximum causes click in the gate following the target gate. This may be explained by trapped charge carriers during the onset of an interrupted avalanche in the target gate. These trapped charges with relatively short lifetime of nanoseconds sometimes cause the avalanche when the voltage rises in the following gate \cite{koehler-sidki2019}. The second observation is the decrease of the full-width at half-maximum (FWHM) of the distributions with increase of the optical pulse's energy.

The click time shift is most prominent at $1.2$ and $1.6$-$\nano\second$ trigger pulse shifts. The origin of this effect is likely electrical, owing to a faster avalanche onset from a multi-photon pulse. The avalanche current rises faster, crosses a fixed threshold of the discriminator inside the SPD earlier and with less timing jitter.

\subsection{Memory effect}
\label{sec:memory}

While conducting the measurements, we have discovered that both detection efficiency and click time shift strongly depend on the history of clicks and pulsed non-blinding illumination applied at the SPD. We call this a memory effect. This effect is distinct from and has a longer timescale than afterpulsing and recovery from deadtime (the latter may also affect these parameters in so-called twilight zone \cite{ware2007}). While the memory effect is known in principle (e.g.,\ \cite{sauge2011} describes a compensation circuit for it in a commercial Si SPD), it has not been explored as a possible security loophole \cite{raupach2022}. This is a good topic for future study.

In our tests, we have found that setting $f$ too high contaminates the data with artefacts owing to the memory effect. To isolate the superlinearity and energy--time effects, we have restricted our measurements to $f = 1~\kilo\hertz$.

If the measurement is conducted at a higher repetition rate, the memory effect amplifies superlinearity. Eve might exploit such higher-rate regime. At $f = 100~\kilo\hertz$, the value of $S$ reaches $3.01$ for SPD1 and $0.92$ for SPD2. \Cref{eq:old_qber} then yields $\text{QBER} = 10.1\%$, which is dangerously low and may allow the attack.

Note that we have measured detection efficiency and superlinearity of the detector itself, meaning we did not take the bit slot that the detector is clicking at into consideration. This may affect the attack significantly. On the one hand, superlinearity may increase dramatically owing to the energy--time effect: the increase of the pulse's energy may move the click from bit slot $N$ to bit slot $N-1$, making the latter bit slot extremely superlinear. On the other hand, \cref{eq:old_qber} does not account for clicks triggered by Eve landing in a wrong bit slot at Bob. Two effects cause this: the energy--time effect and the presence of delayed detection events. The latter has been utilised to make a faint after-gate attack more noticeable by amending the formula for QBER \cite{koehler-sidki2019}. Clicks and double clicks in the wrong bit slot raise QBER, while increased superlinearity in individual bit slots lowers it. Therefore, a rigorous evaluation of the attack requires the knowledge of bit slot timing and implies analysis of the QKD system as a whole, which is beyond the scope of this work.

Nevertheless, \cref{eq:old_qber} could still be applied if Alice and Bob had bit slots that are several detector gates long. For example, our evaluation of the attack holds for a hypothetical QKD system that has laser repetition rate of $312.5 / 5 = 62.5~\mega\hertz$, i.e.,\ period of $5 \times 3.2 = 16~\nano\second$ (5~gates).

\section{Proposed attacks}
\label{sec:attacks}

We believe that the observed energy--time effect can provide Eve a new tool to control Bob's SPDs. Here we suggest two approaches to constructing her attack, without claiming their optimality.

\subsection{Intermediate-basis attack}
\label{sec:attack-intermediate-basis}

We consider a QKD system that uses a weak coherent pulse source and plain BB84 protocol \cite{bennett1984}, which is practical for short-distance links \cite{kurochkin2024}. We assume Bob uses an active-basis-choice scheme with two detectors.

The non-decoy BB84 protocol may be attacked by a photon-number-splitting (PNS) attack \cite{brassard2000,lutkenhaus2002}. However, to fully implement this attack Eve requires a quantum memory, non-destructive photon number measurement, and lossless quantum channel, which are future technologies. Here we propose an implementable alternative attack that exploits commercially available technology of photon-number-resolving detectors \cite{lorenzo2025,idqsnspd} and the energy--time effect. Note that QRate's QKD system \cite{makarov2024} uses decoy states and this attack is not applicable to it.

Eve cuts the quantum channel and measures Alice's states in an intermediate basis, which is $|I_+\rangle = \cos(\alpha)|H\rangle+\sin(\alpha)|V\rangle$, $|I_-\rangle = \sin(\alpha)|H\rangle-\cos(\alpha)|V\rangle$, where $\alpha=\pi/8$. States $|H\rangle$ and $|V\rangle$ denote horizontal and vertical polarisation of the photon.

When Eve detects a single photon, she does not resend anything. But when she gets a double photon, her detection probabilities for $|H\rangle|H\rangle$ and $|D\rangle|D\rangle$ would be $P(|I_+\rangle|I_+\rangle)=\cos^4(\alpha)=0.7286$, $P(|I_+\rangle|I_-\rangle)=2\sin^2(\alpha)\cos^2(\alpha)=0.25$, $P(|I_-\rangle|I_-\rangle)=\sin^4(\alpha)=0.0214$. For $|V\rangle|V\rangle$ and $|A\rangle|A\rangle$, her probabilities are the opposite. Here $|D\rangle = \left(|H\rangle + |V\rangle\right)/\sqrt{2}$, $|A\rangle = \left(|H\rangle - |V\rangle\right)/\sqrt{2}$. Events with measurement outcomes $|I_{+}\rangle|I_{-}\rangle$ or $|I_{-}\rangle|I_{+}\rangle$ are skipped. Now with the double-click event Eve knows the state (but not the basis) with $P=0.7286/(0.7286+0.0214)=0.9715$, or with less than $3\%$ error.

She then resends the same state as a bright classical pulse. Its energy splits into the two Bob's detectors in $\sin^2(\alpha) : \cos^2(\alpha)=1 : 5.828$ ratio, or with $7.7~\deci\bel$ difference. Both click, but the one receiving higher energy clicks earlier. Eve times her pulse at the end of Bob's bit slot, tampering with timing calibration routines if necessary \cite{jain2011,fei2018}. The earlier detector's click is registered in that bit slot, while the other detector clicks in the next bit slot. Bob accepts the first click into his raw key and discards the second click, to prevent a deadtime attack \cite{weier2011,makarov2024}. For example, pulse parameters denoted with black stars in \cref{fig:reaction-mean} result in the change of click timing by approximately $1.9~\nano\second$ for SPD1 and $1.3~\nano\second$ for SPD2. The respective click time shift distributions are relatively narrow and non-overlapping (\cref{fig:reaction-time}), making this step near-deterministic.

In this attack, Eve steals the entire secret key while causing less than 3\% QBER. While she resends only 75\% of her double-photon events, the efficiency of her inducing clicks at Bob may approach 100\%. Overall the performance of our attack comes close to that assumed in security proofs for the PNS attack (which conservatively consider losses in Bob being part of the channel \cite{gottesman2004,hwang2003,scarani2009}). In a realistic setting of lossy Bob's components, our attack likely outperforms the ideal PNS attack and far outperforms an implementable PNS attack that eavesdrops on three-photon pulses \cite{felix2001}.

We note that this attack applies to QKD systems without decoy states and is therefore limited in scope, as most modern implementations employ decoy-state protocols to mitigate multi-photon vulnerabilities. The purpose of this example is to demonstrate how the energy--time effect can be used to control detector outcomes in the intercept-resend scenario, rather than to claim a broadly applicable practical attack.

\subsection{Tampering with synchronisation and recovery from deadtime}
\label{sec:attack-deadtime}

The following attack targets the industrial-prototype QKD system analysed in \cite{makarov2024}. This is a standard prepare-and-measure QKD system that uses the decoy-state BB84 protocol and polarisation encoding. We assume the system is fully updated with countermeasures suggested in its security analysis \cite{makarov2024}. In particular, it has a four-state Bob and simultaneous deadtime of Bob's detectors implemented in software post-processing. The four-state Bob uses his phase modulator (PM) to not only select his measurement basis but also randomly swap or not swap his detector assignment to bit values. The simultaneous deadtime discards any clicks in Bob's detectors that occur earlier than a set number of bit slots from a previous click. At the same time, the system performs timing calibration of Bob's detectors and PM via the communication channel, leaving in principle these timings to be settable by Eve. With all the other countermeasures in place, the analysis \cite{makarov2024} concludes that allowing Eve to control these timings does not constitute a vulnerability.

\begin{figure}
	\includegraphics{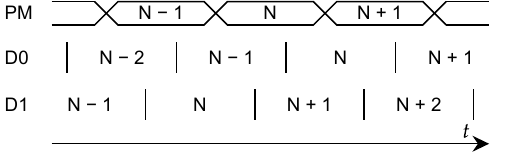}
	\caption{Temporal alignment of bit windows at Bob’s devices needed for the faked-state attack.}
	\label{fig:attack-timing}
\end{figure}

\begin{figure}
	\includegraphics{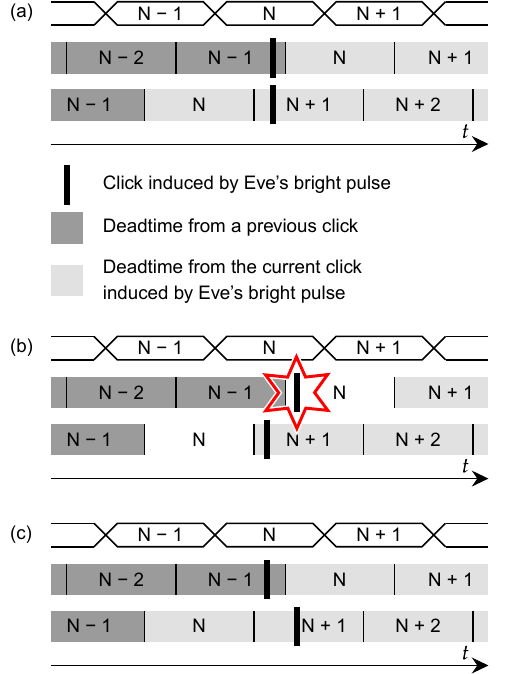}
	\caption{Timing of clicks in Bob under the faked-state attack. (a)~Bob’s basis choice is incompatible with Eve’s. (b)~Bob’s basis choice matches Eve’s and Bob’s assignment of D0 matches Eve’s bit value. (c)~Bob’s basis choice matches Eve’s and Bob’s assignment of D1 matches Eve’s bit value.}
	\label{fig:attack}
\end{figure}

We challenge this by the following attack. Eve begins by tampering with Bob's calibration routines to shift his relative bit windows at his PM and two detectors (D0 and D1) as shown in \cref{fig:attack-timing} \cite{jain2011,fei2018,huang2026}. Here $t$ is time at each component referenced to the passage of a light pulse entering Bob from the communication channel, and $N$ is bit slot number. Then Eve performs an intercept-resend attack, resending a bright light pulse in a state \emph{opposite} that detected by her. E.g.,\ if she detects a vertically-polarised photon from Alice, she resends a horizontally-polarised bright pulse to Bob. She times it to cause clicks in Bob with timing shown in \cref{fig:attack}(a) whenever he chooses a basis incompatible with Eve's basis; the pulse therefore splits evenly between Bob's detectors. She also arranges, by sending an earlier light pulse, that Bob's simultaneous detector deadtime should be lifted from the bit slot $N$ on. However when Bob detects a click in the last slot of his deadtime $N\!-\!1$, he extends his deadtime. This click and the click in slot $N\!+\!1$ are discarded, i.e.,\ they do not contribute to the key.

In the case when Bob chooses the same basis as Eve and he also assigns D0 to the same bit value as Eve's result, Eve's light pulse is almost entirely routed into D1. A minor fraction of it still reaches D0, typically attenuated by $20$--$30~\deci\bel$ owing to the finite quality of Bob's analyser \cite{wang2023}. The timing of click in D0 is thus significantly delayed, landing the click into the bit slot $N$, as shown in \cref{fig:attack}(b). This click is registered in Bob's raw key and matches Eve's bit value. The click in D1 in the slot $N\!+\!1$ is discarded, as it falls into the deadtime.

If Bob assigns D1 to Eve's bit value, the click timing in D0 remains in the slot $N\!-\!1$ [\cref{fig:attack}(c)]. The outcome is then the same as in non-matching basis.

Note that the last click (from Bob's perspective) caused by Eve's faked state is always in D1 in bit slot $N\!+\!1$. It extends Bob's deadtime. Eve can then send another faked state at the end of the deadtime. In case she has failed to detect Alice's photon in that bit slot, she can extend Bob's deadtime further by precisely as much as she needs, via sending a light pulse at a suitable time. She can thus attack the bit slots where she succeeds to detect Alice's photons.

In our experimental data (\cref{fig:reaction-time,fig:reaction-mean}), an energy difference of $25~\deci\bel$ that is typically expected under this attack results in the change of mean click time by as much as $4.1$ ($3.2$)~$\nano\second$ for SPD1 (SPD2), at $1.2$-$\nano\second$ trigger pulse shift. The respective click time distributions are clearly separated, potentially allowing Eve to execute the attack.

Under this attack, Bob is producing a balanced raw key, owing to his random assignment of bit value to D0. If Bob used a standard two-state phase modulation, Eve would only be able to induce one bit value (at least between timing re-calibrations) and her attack would probably fail. Ironically, the four-state Bob is a solution for eliminating other known vulnerabilities \cite{makarov2024}.

While this attack targets the specific implementation \cite{makarov2024}, more generally, attacks of this type rely on how detection events are filtered or discarded, and also on how the calibration procedures are implemented \cite{jain2011,fei2018,huang2026}. For example, clicks may be discarded due to deadtime or because they fall outside valid time windows \cite{weier2011,bourgoin2013,liao2017,li2025}.

\section{Countermeasures}
\label{sec:countermeasures}

First, the detectors should be tested for the presence of the energy--time effect. If found, a possible countermeasure may monitor both detectors clicking within a short time from one another. If such coincident click is registered within the time shift of the energy--time effect, these clicks might be treated similarly to a double-click \cite{lutkenhaus1999,tsurumaru2008}, by assigning them a random bit value. An extended security proof needs to be developed for this countermeasure.

Alternatively, high-energy pulses might be monitored at the entrance of Bob with a separate watchdog detector. However this approach has been criticised as unreliable \cite{lydersen2010a,lydersen2010b,makarov2024} and makes the system hardware more complicated and expensive. The practical effectiveness of such monitoring may depend on its parameters, such as the fraction of light diverted to the watchdog, its bandwidth and sensitivity, and the temporal processing of its output. A practical eavesdropping strategy might then require a prior characterisation of the monitoring system or a tailored temporal pattern of attack pulses.

The photocurrent-monitor countermeasure in the SPD, such as the one used in QRate's QKD system \cite{acheva2023,makarov2024}, may be able to detect multiphoton pulses that induce strong time shift. This needs to be tested.

The particular attack proposed in \cref{sec:attack-deadtime} could be stopped by assigning a random bit value for any click registered in the first bit slot after the deadtime (i.e.,\ treating it in the same way as the double click). As a precaution, QRate has implemented this countermeasure in its QKD system prototype \cite{makarov2024}. In a higher-rate system, treating two or more bit slots in this way may be necessary. However, this countermeasure is not general and does not guarantee the energy--time effect could not be exploited in other unknown attacks.

We remark that measurement-device-independent \cite{lo2012,tang2014} and twin-field \cite{lucamarini2018,wang2018} QKD protocols are immune in principle to detector flaws. Their adoption is another possible solution.

\section{Conclusion}
\label{sec:conclusion}

The prototype sinusoidally-gated SPD by QRate has been tested against the after-gate attack. A limited amount of superlinearity has been found. It remains unclear whether it is sufficient for mounting an attack.

Two unexpected effects have been discovered. In the energy--time effect, the detector click happens earlier under higher optical pulse energy. It starts at hundreds of photons per pulse and continues beyond millions of photons, with our SPD. The total time shift observed is over $2~\nano\second$, however it is not clear whether it has reached saturation or extends further into higher pulse energies. We theoretically propose practical attacks that may exploit this effect. The energy--time effect is not accounted for in the security proofs for QKD. It should be taken into account by theoreticians, engineers, testers, and standards developers. Implementation of the proposed attacks on QKD systems can be a future study.

The memory effect may allow Eve to control SPD parameters at an intermediate time scale. Its significance for QKD security should also be explored.


While this work focuses on gated SPDs, many practical QKD implementations, including terrestrial free-space and satellite-based systems, employ free-running detectors combined with temporal post-selection \cite{bourgoin2013,liao2017,li2025}. In such systems, the pulse-energy-dependent timing shift may similarly affect the assignment of detection events to time windows, potentially introducing additional security considerations. Other detector technologies, such as superconducting-nanowire SPDs \cite{gol'tsman2001,lorenzo2025,idqsnspd}, rely on different physical mechanisms and may exhibit different timing characteristics. These questions can be investigated.

\bigskip

\emph{Acknowledgments:}
We thank QRate for providing the prototype detectors and for discussions. QRate has reviewed this paper before its publication.

\medskip
\emph{Funding:}
Ministry of Science and Education of Russia (program NTI center for quantum communications), Russian Science Foundation (grant 21-42-00040), the Galician Regional Government (consolidation of research units: atlanTTic and own funding through the ``Planes Complementarios de I+D+I con las Comunidades Autonomas'' in Quantum Communication), MICIN with funding from the European Union NextGenerationEU (PRTR-C17.I1), the ``Hub Nacional de Excelencia en Comunicaciones Cu{\' a}nticas'' funded by the Spanish Ministry for Digital Transformation and the Public Service and the European Union NextGenerationEU, and the European Union's Horizon Europe Framework Programme under Marie Sk\l{}odowska-Curie grant 101072637 (project QSI) and project ``Quantum Security Networks Partnership'' (QSNP; grant 101114043).

\medskip
\emph{Author contributions:} The characterisation of superlinearity was developed by all authors. D.K.\ and K.Z.\ discovered the energy--time effect. V.B.\ and V.M.\ noticed the memory effect. The attack in \cref{sec:attack-intermediate-basis} was proposed by K.Z.\ and that in \cref{sec:attack-deadtime} by V.M. D.K.,\ V.B.,\ and K.Z.\ performed the experiment and analysed the data. K.Z.,\ V.B.,\ and V.M.\ wrote the paper with input from all authors. V.M.\ analysed the data and supervised the project.

\medskip
\emph{Data availability:} The data that support the findings of this study are available from the corresponding author upon reasonable request.

\medskip
\emph{Competing interests:} D.K.\ is employed by the vendor that sells the detectors tested. The other authors declare that they have no conflicts of interest and have not received any objections from D.K.\ against reporting the results of this study.

\def\bibsection{\medskip\begin{center}\rule{0.5\columnwidth}{.8pt}\end{center}\medskip} 

\begin{thebibliography}{72}%
\makeatletter
\providecommand \@ifxundefined [1]{%
 \@ifx{#1\undefined}
}%
\providecommand \@ifnum [1]{%
 \ifnum #1\expandafter \@firstoftwo
 \else \expandafter \@secondoftwo
 \fi
}%
\providecommand \@ifx [1]{%
 \ifx #1\expandafter \@firstoftwo
 \else \expandafter \@secondoftwo
 \fi
}%
\providecommand \natexlab [1]{#1}%
\providecommand \enquote  [1]{``#1''}%
\providecommand \bibnamefont  [1]{#1}%
\providecommand \bibfnamefont [1]{#1}%
\providecommand \citenamefont [1]{#1}%
\providecommand \href@noop [0]{\@secondoftwo}%
\providecommand \href [0]{\begingroup \@sanitize@url \@href}%
\providecommand \@href[1]{\@@startlink{#1}\@@href}%
\providecommand \@@href[1]{\endgroup#1\@@endlink}%
\providecommand \@sanitize@url [0]{\catcode `\\12\catcode `\$12\catcode
  `\&12\catcode `\#12\catcode `\^12\catcode `\_12\catcode `\%12\relax}%
\providecommand \@@startlink[1]{}%
\providecommand \@@endlink[0]{}%
\providecommand \url  [0]{\begingroup\@sanitize@url \@url }%
\providecommand \@url [1]{\endgroup\@href {#1}{\urlprefix }}%
\providecommand \urlprefix  [0]{URL }%
\providecommand \Eprint [0]{\href }%
\providecommand \doibase [0]{https://doi.org/}%
\providecommand \selectlanguage [0]{\@gobble}%
\providecommand \bibinfo  [0]{\@secondoftwo}%
\providecommand \bibfield  [0]{\@secondoftwo}%
\providecommand \translation [1]{[#1]}%
\providecommand \BibitemOpen [0]{}%
\providecommand \bibitemStop [0]{}%
\providecommand \bibitemNoStop [0]{.\EOS\space}%
\providecommand \EOS [0]{\spacefactor3000\relax}%
\providecommand \BibitemShut  [1]{\csname bibitem#1\endcsname}%
\let\auto@bib@innerbib\@empty
\bibitem [{\citenamefont {Dowling}\ and\ \citenamefont
  {Milburn}(2003)}]{dowling2003}%
  \BibitemOpen
  \bibfield  {author} {\bibinfo {author} {\bibfnamefont {J.~P.}\ \bibnamefont
  {Dowling}}\ and\ \bibinfo {author} {\bibfnamefont {G.~J.}\ \bibnamefont
  {Milburn}},\ }\bibfield  {title} {\bibinfo {title} {Quantum technology: the
  second quantum revolution},\ }\href {https://doi.org/10.1098/rsta.2003.1227}
  {\bibfield  {journal} {\bibinfo  {journal} {Phil. Trans. R. Soc. Lond. A}\
  }\textbf {\bibinfo {volume} {361}},\ \bibinfo {pages} {1655} (\bibinfo {year}
  {2003})}\BibitemShut {NoStop}%
\bibitem [{\citenamefont {Rivest}\ \emph {et~al.}(1978)\citenamefont {Rivest},
  \citenamefont {Shamir},\ and\ \citenamefont {Adleman}}]{rivest1978}%
  \BibitemOpen
  \bibfield  {author} {\bibinfo {author} {\bibfnamefont {R.~L.}\ \bibnamefont
  {Rivest}}, \bibinfo {author} {\bibfnamefont {A.}~\bibnamefont {Shamir}},\
  and\ \bibinfo {author} {\bibfnamefont {L.}~\bibnamefont {Adleman}},\
  }\bibfield  {title} {\bibinfo {title} {A method for obtaining digital
  signatures and public-key cryptosystems},\ }\href
  {https://doi.org/10.1145/359340.359342} {\bibfield  {journal} {\bibinfo
  {journal} {Commun. ACM}\ }\textbf {\bibinfo {volume} {21}},\ \bibinfo {pages}
  {120} (\bibinfo {year} {1978})}\BibitemShut {NoStop}%
\bibitem [{\citenamefont {Shor}(1997)}]{shor1997}%
  \BibitemOpen
  \bibfield  {author} {\bibinfo {author} {\bibfnamefont {P.~W.}\ \bibnamefont
  {Shor}},\ }\bibfield  {title} {\bibinfo {title} {Polynomial-time algorithms
  for prime factorization and discrete logarithms on a quantum computer},\
  }\href {https://doi.org/10.1137/S0097539795293172} {\bibfield  {journal}
  {\bibinfo  {journal} {SIAM J. Comput.}\ }\textbf {\bibinfo {volume} {26}},\
  \bibinfo {pages} {1484} (\bibinfo {year} {1997})}\BibitemShut {NoStop}%
\bibitem [{\citenamefont {Bennett}\ and\ \citenamefont
  {Brassard}(1984)}]{bennett1984}%
  \BibitemOpen
  \bibfield  {author} {\bibinfo {author} {\bibfnamefont {C.~H.}\ \bibnamefont
  {Bennett}}\ and\ \bibinfo {author} {\bibfnamefont {G.}~\bibnamefont
  {Brassard}},\ }\bibfield  {title} {\bibinfo {title} {Quantum cryptography:
  public key distribution and coin tossing},\ }in\ \href@noop {} {\emph
  {\bibinfo {booktitle} {Proc. International Conference on Computers, Systems,
  and Signal Processing}}}\ (\bibinfo  {publisher} {IEEE Press, New York},\
  \bibinfo {address} {Bangalore, India},\ \bibinfo {year} {1984})\ pp.\
  \bibinfo {pages} {175--179}\BibitemShut {NoStop}%
\bibitem [{\citenamefont {Brassard}\ \emph {et~al.}(2000)\citenamefont
  {Brassard}, \citenamefont {L\"utkenhaus}, \citenamefont {Mor},\ and\
  \citenamefont {Sanders}}]{brassard2000}%
  \BibitemOpen
  \bibfield  {author} {\bibinfo {author} {\bibfnamefont {G.}~\bibnamefont
  {Brassard}}, \bibinfo {author} {\bibfnamefont {N.}~\bibnamefont
  {L\"utkenhaus}}, \bibinfo {author} {\bibfnamefont {T.}~\bibnamefont {Mor}},\
  and\ \bibinfo {author} {\bibfnamefont {B.~C.}\ \bibnamefont {Sanders}},\
  }\bibfield  {title} {\bibinfo {title} {Limitations on practical quantum
  cryptography},\ }\href {https://doi.org/10.1103/PhysRevLett.85.1330}
  {\bibfield  {journal} {\bibinfo  {journal} {Phys. Rev. Lett.}\ }\textbf
  {\bibinfo {volume} {85}},\ \bibinfo {pages} {1330} (\bibinfo {year}
  {2000})}\BibitemShut {NoStop}%
\bibitem [{\citenamefont {Makarov}\ \emph {et~al.}(2006)\citenamefont
  {Makarov}, \citenamefont {Anisimov},\ and\ \citenamefont
  {Skaar}}]{makarov2006}%
  \BibitemOpen
  \bibfield  {author} {\bibinfo {author} {\bibfnamefont {V.}~\bibnamefont
  {Makarov}}, \bibinfo {author} {\bibfnamefont {A.}~\bibnamefont {Anisimov}},\
  and\ \bibinfo {author} {\bibfnamefont {J.}~\bibnamefont {Skaar}},\ }\bibfield
   {title} {\bibinfo {title} {Effects of detector efficiency mismatch on
  security of quantum cryptosystems},\ }\href
  {https://doi.org/10.1103/PhysRevA.74.022313} {\bibfield  {journal} {\bibinfo
  {journal} {Phys. Rev. A}\ }\textbf {\bibinfo {volume} {74}},\ \bibinfo
  {pages} {022313} (\bibinfo {year} {2006})},\ \bibinfo {note} {erratum ibid.
  \textbf{78}, 019905 (2008)}\BibitemShut {NoStop}%
\bibitem [{\citenamefont {Zhao}\ \emph {et~al.}(2008)\citenamefont {Zhao},
  \citenamefont {Fung}, \citenamefont {Qi}, \citenamefont {Chen},\ and\
  \citenamefont {Lo}}]{zhao2008}%
  \BibitemOpen
  \bibfield  {author} {\bibinfo {author} {\bibfnamefont {Y.}~\bibnamefont
  {Zhao}}, \bibinfo {author} {\bibfnamefont {C.-H.~F.}\ \bibnamefont {Fung}},
  \bibinfo {author} {\bibfnamefont {B.}~\bibnamefont {Qi}}, \bibinfo {author}
  {\bibfnamefont {C.}~\bibnamefont {Chen}},\ and\ \bibinfo {author}
  {\bibfnamefont {H.-K.}\ \bibnamefont {Lo}},\ }\bibfield  {title} {\bibinfo
  {title} {Quantum hacking: Experimental demonstration of time-shift attack
  against practical quantum-key-distribution systems},\ }\href
  {https://doi.org/10.1103/PhysRevA.78.042333} {\bibfield  {journal} {\bibinfo
  {journal} {Phys. Rev. A}\ }\textbf {\bibinfo {volume} {78}},\ \bibinfo {eid}
  {042333} (\bibinfo {year} {2008})}\BibitemShut {NoStop}%
\bibitem [{\citenamefont {Xu}\ \emph {et~al.}(2010)\citenamefont {Xu},
  \citenamefont {Qi},\ and\ \citenamefont {Lo}}]{xu2010}%
  \BibitemOpen
  \bibfield  {author} {\bibinfo {author} {\bibfnamefont {F.}~\bibnamefont
  {Xu}}, \bibinfo {author} {\bibfnamefont {B.}~\bibnamefont {Qi}},\ and\
  \bibinfo {author} {\bibfnamefont {H.-K.}\ \bibnamefont {Lo}},\ }\bibfield
  {title} {\bibinfo {title} {Experimental demonstration of phase-remapping
  attack in a practical quantum key distribution system},\ }\href
  {https://doi.org/10.1088/1367-2630/12/11/113026} {\bibfield  {journal}
  {\bibinfo  {journal} {New J. Phys.}\ }\textbf {\bibinfo {volume} {12}},\
  \bibinfo {pages} {113026} (\bibinfo {year} {2010})}\BibitemShut {NoStop}%
\bibitem [{\citenamefont {Lydersen}\ \emph
  {et~al.}(2010{\natexlab{a}})\citenamefont {Lydersen}, \citenamefont
  {Wiechers}, \citenamefont {Wittmann}, \citenamefont {Elser}, \citenamefont
  {Skaar},\ and\ \citenamefont {Makarov}}]{lydersen2010a}%
  \BibitemOpen
  \bibfield  {author} {\bibinfo {author} {\bibfnamefont {L.}~\bibnamefont
  {Lydersen}}, \bibinfo {author} {\bibfnamefont {C.}~\bibnamefont {Wiechers}},
  \bibinfo {author} {\bibfnamefont {C.}~\bibnamefont {Wittmann}}, \bibinfo
  {author} {\bibfnamefont {D.}~\bibnamefont {Elser}}, \bibinfo {author}
  {\bibfnamefont {J.}~\bibnamefont {Skaar}},\ and\ \bibinfo {author}
  {\bibfnamefont {V.}~\bibnamefont {Makarov}},\ }\bibfield  {title} {\bibinfo
  {title} {Hacking commercial quantum cryptography systems by tailored bright
  illumination},\ }\href {https://doi.org/10.1038/nphoton.2010.214} {\bibfield
  {journal} {\bibinfo  {journal} {Nat. Photonics}\ }\textbf {\bibinfo {volume}
  {4}},\ \bibinfo {pages} {686} (\bibinfo {year}
  {2010}{\natexlab{a}})}\BibitemShut {NoStop}%
\bibitem [{\citenamefont {Gerhardt}\ \emph {et~al.}(2011)\citenamefont
  {Gerhardt}, \citenamefont {Liu}, \citenamefont {Lamas-Linares}, \citenamefont
  {Skaar}, \citenamefont {Kurtsiefer},\ and\ \citenamefont
  {Makarov}}]{gerhardt2011}%
  \BibitemOpen
  \bibfield  {author} {\bibinfo {author} {\bibfnamefont {I.}~\bibnamefont
  {Gerhardt}}, \bibinfo {author} {\bibfnamefont {Q.}~\bibnamefont {Liu}},
  \bibinfo {author} {\bibfnamefont {A.}~\bibnamefont {Lamas-Linares}}, \bibinfo
  {author} {\bibfnamefont {J.}~\bibnamefont {Skaar}}, \bibinfo {author}
  {\bibfnamefont {C.}~\bibnamefont {Kurtsiefer}},\ and\ \bibinfo {author}
  {\bibfnamefont {V.}~\bibnamefont {Makarov}},\ }\bibfield  {title} {\bibinfo
  {title} {Full-field implementation of a perfect eavesdropper on a quantum
  cryptography system},\ }\href {https://doi.org/10.1038/ncomms1348} {\bibfield
   {journal} {\bibinfo  {journal} {Nat. Commun.}\ }\textbf {\bibinfo {volume}
  {2}},\ \bibinfo {pages} {349} (\bibinfo {year} {2011})}\BibitemShut {NoStop}%
\bibitem [{\citenamefont {Sun}\ \emph {et~al.}(2011)\citenamefont {Sun},
  \citenamefont {Jiang},\ and\ \citenamefont {Liang}}]{sun2011}%
  \BibitemOpen
  \bibfield  {author} {\bibinfo {author} {\bibfnamefont {S.-H.}\ \bibnamefont
  {Sun}}, \bibinfo {author} {\bibfnamefont {M.-S.}\ \bibnamefont {Jiang}},\
  and\ \bibinfo {author} {\bibfnamefont {L.-M.}\ \bibnamefont {Liang}},\
  }\bibfield  {title} {\bibinfo {title} {Passive {F}araday-mirror attack in a
  practical two-way quantum-key-distribution system},\ }\href
  {https://doi.org/10.1103/PhysRevA.83.062331} {\bibfield  {journal} {\bibinfo
  {journal} {Phys. Rev. A}\ }\textbf {\bibinfo {volume} {83}},\ \bibinfo
  {pages} {062331} (\bibinfo {year} {2011})}\BibitemShut {NoStop}%
\bibitem [{\citenamefont {Jain}\ \emph {et~al.}(2011)\citenamefont {Jain},
  \citenamefont {Wittmann}, \citenamefont {Lydersen}, \citenamefont {Wiechers},
  \citenamefont {Elser}, \citenamefont {Marquardt}, \citenamefont {Makarov},\
  and\ \citenamefont {Leuchs}}]{jain2011}%
  \BibitemOpen
  \bibfield  {author} {\bibinfo {author} {\bibfnamefont {N.}~\bibnamefont
  {Jain}}, \bibinfo {author} {\bibfnamefont {C.}~\bibnamefont {Wittmann}},
  \bibinfo {author} {\bibfnamefont {L.}~\bibnamefont {Lydersen}}, \bibinfo
  {author} {\bibfnamefont {C.}~\bibnamefont {Wiechers}}, \bibinfo {author}
  {\bibfnamefont {D.}~\bibnamefont {Elser}}, \bibinfo {author} {\bibfnamefont
  {C.}~\bibnamefont {Marquardt}}, \bibinfo {author} {\bibfnamefont
  {V.}~\bibnamefont {Makarov}},\ and\ \bibinfo {author} {\bibfnamefont
  {G.}~\bibnamefont {Leuchs}},\ }\bibfield  {title} {\bibinfo {title} {Device
  calibration impacts security of quantum key distribution},\ }\href
  {https://doi.org/10.1103/PhysRevLett.107.110501} {\bibfield  {journal}
  {\bibinfo  {journal} {Phys. Rev. Lett.}\ }\textbf {\bibinfo {volume} {107}},\
  \bibinfo {pages} {110501} (\bibinfo {year} {2011})}\BibitemShut {NoStop}%
\bibitem [{\citenamefont {Weier}\ \emph {et~al.}(2011)\citenamefont {Weier},
  \citenamefont {Krauss}, \citenamefont {Rau}, \citenamefont {F{\"u}rst},
  \citenamefont {Nauerth},\ and\ \citenamefont {Weinfurter}}]{weier2011}%
  \BibitemOpen
  \bibfield  {author} {\bibinfo {author} {\bibfnamefont {H.}~\bibnamefont
  {Weier}}, \bibinfo {author} {\bibfnamefont {H.}~\bibnamefont {Krauss}},
  \bibinfo {author} {\bibfnamefont {M.}~\bibnamefont {Rau}}, \bibinfo {author}
  {\bibfnamefont {M.}~\bibnamefont {F{\"u}rst}}, \bibinfo {author}
  {\bibfnamefont {S.}~\bibnamefont {Nauerth}},\ and\ \bibinfo {author}
  {\bibfnamefont {H.}~\bibnamefont {Weinfurter}},\ }\bibfield  {title}
  {\bibinfo {title} {Quantum eavesdropping without interception: an attack
  exploiting the dead time of single-photon detectors},\ }\href
  {https://doi.org/10.1088/1367-2630/13/7/073024} {\bibfield  {journal}
  {\bibinfo  {journal} {New J. Phys.}\ }\textbf {\bibinfo {volume} {13}},\
  \bibinfo {pages} {073024} (\bibinfo {year} {2011})}\BibitemShut {NoStop}%
\bibitem [{\citenamefont {Li}\ \emph {et~al.}(2011)\citenamefont {Li},
  \citenamefont {Wang}, \citenamefont {Huang}, \citenamefont {Chen},
  \citenamefont {Yin}, \citenamefont {Li}, \citenamefont {Zhou}, \citenamefont
  {Liu}, \citenamefont {Zhang}, \citenamefont {Guo}, \citenamefont {Bao},\ and\
  \citenamefont {Han}}]{li2011a}%
  \BibitemOpen
  \bibfield  {author} {\bibinfo {author} {\bibfnamefont {H.-W.}\ \bibnamefont
  {Li}}, \bibinfo {author} {\bibfnamefont {S.}~\bibnamefont {Wang}}, \bibinfo
  {author} {\bibfnamefont {J.-Z.}\ \bibnamefont {Huang}}, \bibinfo {author}
  {\bibfnamefont {W.}~\bibnamefont {Chen}}, \bibinfo {author} {\bibfnamefont
  {Z.-Q.}\ \bibnamefont {Yin}}, \bibinfo {author} {\bibfnamefont {F.-Y.}\
  \bibnamefont {Li}}, \bibinfo {author} {\bibfnamefont {Z.}~\bibnamefont
  {Zhou}}, \bibinfo {author} {\bibfnamefont {D.}~\bibnamefont {Liu}}, \bibinfo
  {author} {\bibfnamefont {Y.}~\bibnamefont {Zhang}}, \bibinfo {author}
  {\bibfnamefont {G.-C.}\ \bibnamefont {Guo}}, \bibinfo {author} {\bibfnamefont
  {W.-S.}\ \bibnamefont {Bao}},\ and\ \bibinfo {author} {\bibfnamefont {Z.-F.}\
  \bibnamefont {Han}},\ }\bibfield  {title} {\bibinfo {title} {Attacking a
  practical quantum-key-distribution system with wavelength-dependent
  beam-splitter and multiwavelength sources},\ }\href
  {https://doi.org/10.1103/PhysRevA.84.062308} {\bibfield  {journal} {\bibinfo
  {journal} {Phys. Rev. A}\ }\textbf {\bibinfo {volume} {84}},\ \bibinfo
  {pages} {062308} (\bibinfo {year} {2011})}\BibitemShut {NoStop}%
\bibitem [{\citenamefont {Jiang}\ \emph {et~al.}(2012)\citenamefont {Jiang},
  \citenamefont {Sun}, \citenamefont {Li},\ and\ \citenamefont
  {Liang}}]{jiang2012}%
  \BibitemOpen
  \bibfield  {author} {\bibinfo {author} {\bibfnamefont {M.-S.}\ \bibnamefont
  {Jiang}}, \bibinfo {author} {\bibfnamefont {S.-H.}\ \bibnamefont {Sun}},
  \bibinfo {author} {\bibfnamefont {C.-Y.}\ \bibnamefont {Li}},\ and\ \bibinfo
  {author} {\bibfnamefont {L.-M.}\ \bibnamefont {Liang}},\ }\bibfield  {title}
  {\bibinfo {title} {Wavelength-selected photon-number-splitting attack against
  plug-and-play quantum key distribution systems with decoy states},\ }\href
  {https://doi.org/10.1103/PhysRevA.86.032310} {\bibfield  {journal} {\bibinfo
  {journal} {Phys. Rev. A}\ }\textbf {\bibinfo {volume} {86}},\ \bibinfo
  {pages} {032310} (\bibinfo {year} {2012})}\BibitemShut {NoStop}%
\bibitem [{\citenamefont {Jouguet}\ \emph {et~al.}(2013)\citenamefont
  {Jouguet}, \citenamefont {Kunz-Jacques},\ and\ \citenamefont
  {Diamanti}}]{jouguet2013}%
  \BibitemOpen
  \bibfield  {author} {\bibinfo {author} {\bibfnamefont {P.}~\bibnamefont
  {Jouguet}}, \bibinfo {author} {\bibfnamefont {S.}~\bibnamefont
  {Kunz-Jacques}},\ and\ \bibinfo {author} {\bibfnamefont {E.}~\bibnamefont
  {Diamanti}},\ }\bibfield  {title} {\bibinfo {title} {Preventing calibration
  attacks on the local oscillator in continuous-variable quantum key
  distribution},\ }\href {https://doi.org/10.1103/PhysRevA.87.062313}
  {\bibfield  {journal} {\bibinfo  {journal} {Phys. Rev. A}\ }\textbf {\bibinfo
  {volume} {87}},\ \bibinfo {pages} {062313} (\bibinfo {year}
  {2013})}\BibitemShut {NoStop}%
\bibitem [{\citenamefont {Jain}\ \emph {et~al.}(2014)\citenamefont {Jain},
  \citenamefont {Anisimova}, \citenamefont {Khan}, \citenamefont {Makarov},
  \citenamefont {Marquardt},\ and\ \citenamefont {Leuchs}}]{jain2014}%
  \BibitemOpen
  \bibfield  {author} {\bibinfo {author} {\bibfnamefont {N.}~\bibnamefont
  {Jain}}, \bibinfo {author} {\bibfnamefont {E.}~\bibnamefont {Anisimova}},
  \bibinfo {author} {\bibfnamefont {I.}~\bibnamefont {Khan}}, \bibinfo {author}
  {\bibfnamefont {V.}~\bibnamefont {Makarov}}, \bibinfo {author} {\bibfnamefont
  {C.}~\bibnamefont {Marquardt}},\ and\ \bibinfo {author} {\bibfnamefont
  {G.}~\bibnamefont {Leuchs}},\ }\bibfield  {title} {\bibinfo {title}
  {Trojan-horse attacks threaten the security of practical quantum
  cryptography},\ }\href {https://doi.org/10.1088/1367-2630/16/12/123030}
  {\bibfield  {journal} {\bibinfo  {journal} {New J. Phys.}\ }\textbf {\bibinfo
  {volume} {16}},\ \bibinfo {pages} {123030} (\bibinfo {year}
  {2014})}\BibitemShut {NoStop}%
\bibitem [{\citenamefont {Sajeed}\ \emph
  {et~al.}(2015{\natexlab{a}})\citenamefont {Sajeed}, \citenamefont
  {Radchenko}, \citenamefont {Kaiser}, \citenamefont {Bourgoin}, \citenamefont
  {Pappa}, \citenamefont {Monat}, \citenamefont {Legr\'e},\ and\ \citenamefont
  {Makarov}}]{sajeed2015}%
  \BibitemOpen
  \bibfield  {author} {\bibinfo {author} {\bibfnamefont {S.}~\bibnamefont
  {Sajeed}}, \bibinfo {author} {\bibfnamefont {I.}~\bibnamefont {Radchenko}},
  \bibinfo {author} {\bibfnamefont {S.}~\bibnamefont {Kaiser}}, \bibinfo
  {author} {\bibfnamefont {J.-P.}\ \bibnamefont {Bourgoin}}, \bibinfo {author}
  {\bibfnamefont {A.}~\bibnamefont {Pappa}}, \bibinfo {author} {\bibfnamefont
  {L.}~\bibnamefont {Monat}}, \bibinfo {author} {\bibfnamefont
  {M.}~\bibnamefont {Legr\'e}},\ and\ \bibinfo {author} {\bibfnamefont
  {V.}~\bibnamefont {Makarov}},\ }\bibfield  {title} {\bibinfo {title} {Attacks
  exploiting deviation of mean photon number in quantum key distribution and
  coin tossing},\ }\href {https://doi.org/10.1103/PhysRevA.91.032326}
  {\bibfield  {journal} {\bibinfo  {journal} {Phys. Rev. A}\ }\textbf {\bibinfo
  {volume} {91}},\ \bibinfo {pages} {032326} (\bibinfo {year}
  {2015}{\natexlab{a}})}\BibitemShut {NoStop}%
\bibitem [{\citenamefont {Rau}\ \emph {et~al.}(2015)\citenamefont {Rau},
  \citenamefont {Vogl}, \citenamefont {Corrielli}, \citenamefont {Vest},
  \citenamefont {Fuchs}, \citenamefont {Nauerth},\ and\ \citenamefont
  {Weinfurter}}]{rau2015}%
  \BibitemOpen
  \bibfield  {author} {\bibinfo {author} {\bibfnamefont {M.}~\bibnamefont
  {Rau}}, \bibinfo {author} {\bibfnamefont {T.}~\bibnamefont {Vogl}}, \bibinfo
  {author} {\bibfnamefont {G.}~\bibnamefont {Corrielli}}, \bibinfo {author}
  {\bibfnamefont {G.}~\bibnamefont {Vest}}, \bibinfo {author} {\bibfnamefont
  {L.}~\bibnamefont {Fuchs}}, \bibinfo {author} {\bibfnamefont
  {S.}~\bibnamefont {Nauerth}},\ and\ \bibinfo {author} {\bibfnamefont
  {H.}~\bibnamefont {Weinfurter}},\ }\bibfield  {title} {\bibinfo {title}
  {Spatial mode side channels in free-space {QKD} implementations},\ }\href
  {https://doi.org/10.1109/JSTQE.2014.2372008} {\bibfield  {journal} {\bibinfo
  {journal} {IEEE J. Quantum. Electron.}\ }\textbf {\bibinfo {volume} {21}},\
  \bibinfo {pages} {6600905} (\bibinfo {year} {2015})}\BibitemShut {NoStop}%
\bibitem [{\citenamefont {Sajeed}\ \emph
  {et~al.}(2015{\natexlab{b}})\citenamefont {Sajeed}, \citenamefont
  {Chaiwongkhot}, \citenamefont {Bourgoin}, \citenamefont {Jennewein},
  \citenamefont {L{\" u}tkenhaus},\ and\ \citenamefont
  {Makarov}}]{sajeed2015a}%
  \BibitemOpen
  \bibfield  {author} {\bibinfo {author} {\bibfnamefont {S.}~\bibnamefont
  {Sajeed}}, \bibinfo {author} {\bibfnamefont {P.}~\bibnamefont
  {Chaiwongkhot}}, \bibinfo {author} {\bibfnamefont {J.-P.}\ \bibnamefont
  {Bourgoin}}, \bibinfo {author} {\bibfnamefont {T.}~\bibnamefont {Jennewein}},
  \bibinfo {author} {\bibfnamefont {N.}~\bibnamefont {L{\" u}tkenhaus}},\ and\
  \bibinfo {author} {\bibfnamefont {V.}~\bibnamefont {Makarov}},\ }\bibfield
  {title} {\bibinfo {title} {Security loophole in free-space quantum key
  distribution due to spatial-mode detector-efficiency mismatch},\ }\href
  {https://doi.org/10.1103/PhysRevA.91.062301} {\bibfield  {journal} {\bibinfo
  {journal} {Phys. Rev. A}\ }\textbf {\bibinfo {volume} {91}},\ \bibinfo
  {pages} {062301} (\bibinfo {year} {2015}{\natexlab{b}})}\BibitemShut
  {NoStop}%
\bibitem [{\citenamefont {Makarov}\ \emph {et~al.}(2016)\citenamefont
  {Makarov}, \citenamefont {Bourgoin}, \citenamefont {Chaiwongkhot},
  \citenamefont {Gagn{\'e}}, \citenamefont {Jennewein}, \citenamefont {Kaiser},
  \citenamefont {Kashyap}, \citenamefont {Legr{\'e}}, \citenamefont
  {Minshull},\ and\ \citenamefont {Sajeed}}]{makarov2016}%
  \BibitemOpen
  \bibfield  {author} {\bibinfo {author} {\bibfnamefont {V.}~\bibnamefont
  {Makarov}}, \bibinfo {author} {\bibfnamefont {J.-P.}\ \bibnamefont
  {Bourgoin}}, \bibinfo {author} {\bibfnamefont {P.}~\bibnamefont
  {Chaiwongkhot}}, \bibinfo {author} {\bibfnamefont {M.}~\bibnamefont
  {Gagn{\'e}}}, \bibinfo {author} {\bibfnamefont {T.}~\bibnamefont
  {Jennewein}}, \bibinfo {author} {\bibfnamefont {S.}~\bibnamefont {Kaiser}},
  \bibinfo {author} {\bibfnamefont {R.}~\bibnamefont {Kashyap}}, \bibinfo
  {author} {\bibfnamefont {M.}~\bibnamefont {Legr{\'e}}}, \bibinfo {author}
  {\bibfnamefont {C.}~\bibnamefont {Minshull}},\ and\ \bibinfo {author}
  {\bibfnamefont {S.}~\bibnamefont {Sajeed}},\ }\bibfield  {title} {\bibinfo
  {title} {Creation of backdoors in quantum communications via laser damage},\
  }\href {https://doi.org/10.1103/PhysRevA.94.030302} {\bibfield  {journal}
  {\bibinfo  {journal} {Phys. Rev. A}\ }\textbf {\bibinfo {volume} {94}},\
  \bibinfo {pages} {030302} (\bibinfo {year} {2016})}\BibitemShut {NoStop}%
\bibitem [{\citenamefont {Yoshino}\ \emph {et~al.}(2018)\citenamefont
  {Yoshino}, \citenamefont {Fujiwara}, \citenamefont {Nakata}, \citenamefont
  {Sumiya}, \citenamefont {Sasaki}, \citenamefont {Takeoka}, \citenamefont
  {Sasaki}, \citenamefont {Tajima}, \citenamefont {Koashi},\ and\ \citenamefont
  {Tomita}}]{yoshino2018}%
  \BibitemOpen
  \bibfield  {author} {\bibinfo {author} {\bibfnamefont {K.~I.}\ \bibnamefont
  {Yoshino}}, \bibinfo {author} {\bibfnamefont {M.}~\bibnamefont {Fujiwara}},
  \bibinfo {author} {\bibfnamefont {K.}~\bibnamefont {Nakata}}, \bibinfo
  {author} {\bibfnamefont {T.}~\bibnamefont {Sumiya}}, \bibinfo {author}
  {\bibfnamefont {T.}~\bibnamefont {Sasaki}}, \bibinfo {author} {\bibfnamefont
  {M.}~\bibnamefont {Takeoka}}, \bibinfo {author} {\bibfnamefont
  {M.}~\bibnamefont {Sasaki}}, \bibinfo {author} {\bibfnamefont
  {A.}~\bibnamefont {Tajima}}, \bibinfo {author} {\bibfnamefont
  {M.}~\bibnamefont {Koashi}},\ and\ \bibinfo {author} {\bibfnamefont
  {A.}~\bibnamefont {Tomita}},\ }\bibfield  {title} {\bibinfo {title} {Quantum
  key distribution with an efficient countermeasure against correlated
  intensity fluctuations in optical pulses},\ }\href
  {https://doi.org/10.1038/s41534-017-0057-8} {\bibfield  {journal} {\bibinfo
  {journal} {npj Quantum Inf.}\ }\textbf {\bibinfo {volume} {4}},\ \bibinfo
  {pages} {8} (\bibinfo {year} {2018})}\BibitemShut {NoStop}%
\bibitem [{\citenamefont {Huang}\ \emph {et~al.}(2018)\citenamefont {Huang},
  \citenamefont {Sun}, \citenamefont {Liu},\ and\ \citenamefont
  {Makarov}}]{huang2018}%
  \BibitemOpen
  \bibfield  {author} {\bibinfo {author} {\bibfnamefont {A.}~\bibnamefont
  {Huang}}, \bibinfo {author} {\bibfnamefont {S.~H.}\ \bibnamefont {Sun}},
  \bibinfo {author} {\bibfnamefont {Z.}~\bibnamefont {Liu}},\ and\ \bibinfo
  {author} {\bibfnamefont {V.}~\bibnamefont {Makarov}},\ }\bibfield  {title}
  {\bibinfo {title} {Quantum key distribution with distinguishable decoy
  states},\ }\href {https://doi.org/10.1103/PhysRevA.98.012330} {\bibfield
  {journal} {\bibinfo  {journal} {Phys. Rev. A}\ }\textbf {\bibinfo {volume}
  {98}},\ \bibinfo {pages} {012330} (\bibinfo {year} {2018})}\BibitemShut
  {NoStop}%
\bibitem [{\citenamefont {Huang}\ \emph {et~al.}(2019)\citenamefont {Huang},
  \citenamefont {Navarrete}, \citenamefont {Sun}, \citenamefont {Chaiwongkhot},
  \citenamefont {Curty},\ and\ \citenamefont {Makarov}}]{huang2019}%
  \BibitemOpen
  \bibfield  {author} {\bibinfo {author} {\bibfnamefont {A.}~\bibnamefont
  {Huang}}, \bibinfo {author} {\bibfnamefont {{\'A}.}~\bibnamefont
  {Navarrete}}, \bibinfo {author} {\bibfnamefont {S.-H.}\ \bibnamefont {Sun}},
  \bibinfo {author} {\bibfnamefont {P.}~\bibnamefont {Chaiwongkhot}}, \bibinfo
  {author} {\bibfnamefont {M.}~\bibnamefont {Curty}},\ and\ \bibinfo {author}
  {\bibfnamefont {V.}~\bibnamefont {Makarov}},\ }\bibfield  {title} {\bibinfo
  {title} {Laser-seeding attack in quantum key distribution},\ }\href
  {https://doi.org/10.1103/PhysRevApplied.12.064043} {\bibfield  {journal}
  {\bibinfo  {journal} {Phys. Rev. Appl.}\ }\textbf {\bibinfo {volume} {12}},\
  \bibinfo {pages} {064043} (\bibinfo {year} {2019})}\BibitemShut {NoStop}%
\bibitem [{\citenamefont {Huang}\ \emph {et~al.}(2023)\citenamefont {Huang},
  \citenamefont {Mizutani}, \citenamefont {Lo}, \citenamefont {Makarov},\ and\
  \citenamefont {Tamaki}}]{huang2023}%
  \BibitemOpen
  \bibfield  {author} {\bibinfo {author} {\bibfnamefont {A.}~\bibnamefont
  {Huang}}, \bibinfo {author} {\bibfnamefont {A.}~\bibnamefont {Mizutani}},
  \bibinfo {author} {\bibfnamefont {H.-K.}\ \bibnamefont {Lo}}, \bibinfo
  {author} {\bibfnamefont {V.}~\bibnamefont {Makarov}},\ and\ \bibinfo {author}
  {\bibfnamefont {K.}~\bibnamefont {Tamaki}},\ }\bibfield  {title} {\bibinfo
  {title} {Characterization of state-preparation uncertainty in quantum key
  distribution},\ }\href {https://doi.org/10.1103/PhysRevApplied.19.014048}
  {\bibfield  {journal} {\bibinfo  {journal} {Phys. Rev. Lett.}\ }\textbf
  {\bibinfo {volume} {19}},\ \bibinfo {pages} {014048} (\bibinfo {year}
  {2023})}\BibitemShut {NoStop}%
\bibitem [{\citenamefont {Ye}\ \emph {et~al.}(2023)\citenamefont {Ye},
  \citenamefont {Chen}, \citenamefont {Zhang}, \citenamefont {Lu},
  \citenamefont {Wang}, \citenamefont {Huang}, \citenamefont {Wang},
  \citenamefont {He}, \citenamefont {Yin}, \citenamefont {Guo},\ and\
  \citenamefont {Han}}]{ye2023}%
  \BibitemOpen
  \bibfield  {author} {\bibinfo {author} {\bibfnamefont {P.}~\bibnamefont
  {Ye}}, \bibinfo {author} {\bibfnamefont {W.}~\bibnamefont {Chen}}, \bibinfo
  {author} {\bibfnamefont {G.-W.}\ \bibnamefont {Zhang}}, \bibinfo {author}
  {\bibfnamefont {F.-Y.}\ \bibnamefont {Lu}}, \bibinfo {author} {\bibfnamefont
  {F.-X.}\ \bibnamefont {Wang}}, \bibinfo {author} {\bibfnamefont {G.-Z.}\
  \bibnamefont {Huang}}, \bibinfo {author} {\bibfnamefont {S.}~\bibnamefont
  {Wang}}, \bibinfo {author} {\bibfnamefont {D.-Y.}\ \bibnamefont {He}},
  \bibinfo {author} {\bibfnamefont {Z.-Q.}\ \bibnamefont {Yin}}, \bibinfo
  {author} {\bibfnamefont {G.-C.}\ \bibnamefont {Guo}},\ and\ \bibinfo {author}
  {\bibfnamefont {Z.-F.}\ \bibnamefont {Han}},\ }\bibfield  {title} {\bibinfo
  {title} {Induced-photorefraction attack against quantum key distribution},\
  }\href {https://doi.org/10.1103/PhysRevApplied.19.054052} {\bibfield
  {journal} {\bibinfo  {journal} {Phys. Rev. Appl.}\ }\textbf {\bibinfo
  {volume} {19}},\ \bibinfo {pages} {054052} (\bibinfo {year}
  {2023})}\BibitemShut {NoStop}%
\bibitem [{\citenamefont {Lu}\ \emph {et~al.}(2023)\citenamefont {Lu},
  \citenamefont {Ye}, \citenamefont {Wang}, \citenamefont {Wang}, \citenamefont
  {Yin}, \citenamefont {Wang}, \citenamefont {Huang}, \citenamefont {Chen},
  \citenamefont {He}, \citenamefont {Fan-Yuan}, \citenamefont {Guo},\ and\
  \citenamefont {Han}}]{lu2023}%
  \BibitemOpen
  \bibfield  {author} {\bibinfo {author} {\bibfnamefont {F.-Y.}\ \bibnamefont
  {Lu}}, \bibinfo {author} {\bibfnamefont {P.}~\bibnamefont {Ye}}, \bibinfo
  {author} {\bibfnamefont {Z.-H.}\ \bibnamefont {Wang}}, \bibinfo {author}
  {\bibfnamefont {S.}~\bibnamefont {Wang}}, \bibinfo {author} {\bibfnamefont
  {Z.-Q.}\ \bibnamefont {Yin}}, \bibinfo {author} {\bibfnamefont
  {R.}~\bibnamefont {Wang}}, \bibinfo {author} {\bibfnamefont {X.-J.}\
  \bibnamefont {Huang}}, \bibinfo {author} {\bibfnamefont {W.}~\bibnamefont
  {Chen}}, \bibinfo {author} {\bibfnamefont {D.-Y.}\ \bibnamefont {He}},
  \bibinfo {author} {\bibfnamefont {G.-J.}\ \bibnamefont {Fan-Yuan}}, \bibinfo
  {author} {\bibfnamefont {G.-C.}\ \bibnamefont {Guo}},\ and\ \bibinfo {author}
  {\bibfnamefont {Z.-F.}\ \bibnamefont {Han}},\ }\bibfield  {title} {\bibinfo
  {title} {Hacking measurement-device-independent quantum key distribution},\
  }\href {https://doi.org/10.1364/OPTICA.485389} {\bibfield  {journal}
  {\bibinfo  {journal} {Optica}\ }\textbf {\bibinfo {volume} {10}},\ \bibinfo
  {pages} {520} (\bibinfo {year} {2023})}\BibitemShut {NoStop}%
\bibitem [{\citenamefont {Baliuka}\ \emph {et~al.}(2023)\citenamefont
  {Baliuka}, \citenamefont {St{\"o}cker}, \citenamefont {Auer}, \citenamefont
  {Freiwang}, \citenamefont {Weinfurter},\ and\ \citenamefont
  {Knips}}]{baliuka2023}%
  \BibitemOpen
  \bibfield  {author} {\bibinfo {author} {\bibfnamefont {A.}~\bibnamefont
  {Baliuka}}, \bibinfo {author} {\bibfnamefont {M.}~\bibnamefont
  {St{\"o}cker}}, \bibinfo {author} {\bibfnamefont {M.}~\bibnamefont {Auer}},
  \bibinfo {author} {\bibfnamefont {P.}~\bibnamefont {Freiwang}}, \bibinfo
  {author} {\bibfnamefont {H.}~\bibnamefont {Weinfurter}},\ and\ \bibinfo
  {author} {\bibfnamefont {L.}~\bibnamefont {Knips}},\ }\bibfield  {title}
  {\bibinfo {title} {Deep-learning-based radio-frequency side-channel attack on
  quantum key distribution},\ }\href
  {https://doi.org/10.1103/PhysRevApplied.20.054040} {\bibfield  {journal}
  {\bibinfo  {journal} {Phys. Rev. Appl.}\ }\textbf {\bibinfo {volume} {20}},\
  \bibinfo {pages} {054040} (\bibinfo {year} {2023})}\BibitemShut {NoStop}%
\bibitem [{\citenamefont {Ac\'{\i}n}\ \emph {et~al.}(2006)\citenamefont
  {Ac\'{\i}n}, \citenamefont {Gisin},\ and\ \citenamefont
  {Masanes}}]{acin2006}%
  \BibitemOpen
  \bibfield  {author} {\bibinfo {author} {\bibfnamefont {A.}~\bibnamefont
  {Ac\'{\i}n}}, \bibinfo {author} {\bibfnamefont {N.}~\bibnamefont {Gisin}},\
  and\ \bibinfo {author} {\bibfnamefont {L.}~\bibnamefont {Masanes}},\
  }\bibfield  {title} {\bibinfo {title} {From {B}ell's theorem to secure
  quantum key distribution},\ }\href
  {https://doi.org/10.1103/PhysRevLett.97.120405} {\bibfield  {journal}
  {\bibinfo  {journal} {Phys. Rev. Lett.}\ }\textbf {\bibinfo {volume} {97}},\
  \bibinfo {pages} {120405} (\bibinfo {year} {2006})}\BibitemShut {NoStop}%
\bibitem [{\citenamefont {Briegel}\ \emph {et~al.}(1998)\citenamefont
  {Briegel}, \citenamefont {D\"ur}, \citenamefont {Cirac},\ and\ \citenamefont
  {Zoller}}]{briegel1998}%
  \BibitemOpen
  \bibfield  {author} {\bibinfo {author} {\bibfnamefont {H.-J.}\ \bibnamefont
  {Briegel}}, \bibinfo {author} {\bibfnamefont {W.}~\bibnamefont {D\"ur}},
  \bibinfo {author} {\bibfnamefont {J.~I.}\ \bibnamefont {Cirac}},\ and\
  \bibinfo {author} {\bibfnamefont {P.}~\bibnamefont {Zoller}},\ }\bibfield
  {title} {\bibinfo {title} {Quantum repeaters: The role of imperfect local
  operations in quantum communication},\ }\href
  {https://doi.org/10.1103/PhysRevLett.81.5932} {\bibfield  {journal} {\bibinfo
   {journal} {Phys. Rev. Lett.}\ }\textbf {\bibinfo {volume} {81}},\ \bibinfo
  {pages} {5932} (\bibinfo {year} {1998})}\BibitemShut {NoStop}%
\bibitem [{\citenamefont {Lo}\ \emph {et~al.}(2012)\citenamefont {Lo},
  \citenamefont {Curty},\ and\ \citenamefont {Qi}}]{lo2012}%
  \BibitemOpen
  \bibfield  {author} {\bibinfo {author} {\bibfnamefont {H.-K.}\ \bibnamefont
  {Lo}}, \bibinfo {author} {\bibfnamefont {M.}~\bibnamefont {Curty}},\ and\
  \bibinfo {author} {\bibfnamefont {B.}~\bibnamefont {Qi}},\ }\bibfield
  {title} {\bibinfo {title} {Measurement-device-independent quantum key
  distribution},\ }\href {https://doi.org/10.1103/PhysRevLett.108.130503}
  {\bibfield  {journal} {\bibinfo  {journal} {Phys. Rev. Lett.}\ }\textbf
  {\bibinfo {volume} {108}},\ \bibinfo {pages} {130503} (\bibinfo {year}
  {2012})}\BibitemShut {NoStop}%
\bibitem [{\citenamefont {Tang}\ \emph {et~al.}(2014)\citenamefont {Tang},
  \citenamefont {Liao}, \citenamefont {Xu}, \citenamefont {Qi}, \citenamefont
  {Qian},\ and\ \citenamefont {Lo}}]{tang2014}%
  \BibitemOpen
  \bibfield  {author} {\bibinfo {author} {\bibfnamefont {Z.}~\bibnamefont
  {Tang}}, \bibinfo {author} {\bibfnamefont {Z.}~\bibnamefont {Liao}}, \bibinfo
  {author} {\bibfnamefont {F.}~\bibnamefont {Xu}}, \bibinfo {author}
  {\bibfnamefont {B.}~\bibnamefont {Qi}}, \bibinfo {author} {\bibfnamefont
  {L.}~\bibnamefont {Qian}},\ and\ \bibinfo {author} {\bibfnamefont {H.-K.}\
  \bibnamefont {Lo}},\ }\bibfield  {title} {\bibinfo {title} {Experimental
  demonstration of polarization encoding measurement-device-independent quantum
  key distribution},\ }\href {https://doi.org/10.1103/PhysRevLett.112.190503}
  {\bibfield  {journal} {\bibinfo  {journal} {Phys. Rev. Lett.}\ }\textbf
  {\bibinfo {volume} {112}},\ \bibinfo {pages} {190503} (\bibinfo {year}
  {2014})}\BibitemShut {NoStop}%
\bibitem [{\citenamefont {Ma}\ \emph {et~al.}(2007)\citenamefont {Ma},
  \citenamefont {Fung},\ and\ \citenamefont {Lo}}]{xiongfeng2007}%
  \BibitemOpen
  \bibfield  {author} {\bibinfo {author} {\bibfnamefont {X.}~\bibnamefont
  {Ma}}, \bibinfo {author} {\bibfnamefont {C.-H.~F.}\ \bibnamefont {Fung}},\
  and\ \bibinfo {author} {\bibfnamefont {H.-K.}\ \bibnamefont {Lo}},\
  }\bibfield  {title} {\bibinfo {title} {Quantum key distribution with
  entangled photon sources},\ }\href
  {https://doi.org/10.1103/PhysRevA.76.012307} {\bibfield  {journal} {\bibinfo
  {journal} {Phys. Rev. A}\ }\textbf {\bibinfo {volume} {76}},\ \bibinfo
  {pages} {012307} (\bibinfo {year} {2007})}\BibitemShut {NoStop}%
\bibitem [{\citenamefont {Cao}\ \emph {et~al.}(2016)\citenamefont {Cao},
  \citenamefont {Zhou}, \citenamefont {Yuan},\ and\ \citenamefont
  {Ma}}]{zhu2016}%
  \BibitemOpen
  \bibfield  {author} {\bibinfo {author} {\bibfnamefont {Z.}~\bibnamefont
  {Cao}}, \bibinfo {author} {\bibfnamefont {H.}~\bibnamefont {Zhou}}, \bibinfo
  {author} {\bibfnamefont {X.}~\bibnamefont {Yuan}},\ and\ \bibinfo {author}
  {\bibfnamefont {X.}~\bibnamefont {Ma}},\ }\bibfield  {title} {\bibinfo
  {title} {Source-independent quantum random number generation},\ }\href
  {https://doi.org/10.1103/PhysRevX.6.011020} {\bibfield  {journal} {\bibinfo
  {journal} {Phys. Rev. X}\ }\textbf {\bibinfo {volume} {6}},\ \bibinfo {pages}
  {011020} (\bibinfo {year} {2016})}\BibitemShut {NoStop}%
\bibitem [{\citenamefont {Gottesman}\ and\ \citenamefont
  {Chuang}(1999)}]{gottesman1999}%
  \BibitemOpen
  \bibfield  {author} {\bibinfo {author} {\bibfnamefont {D.}~\bibnamefont
  {Gottesman}}\ and\ \bibinfo {author} {\bibfnamefont {I.}~\bibnamefont
  {Chuang}},\ }\bibfield  {title} {\bibinfo {title} {Demonstrating the
  viability of universal quantum computation using teleportation and
  single-qubit operations.},\ }\href {https://doi.org/10.1038/46503} {\bibfield
   {journal} {\bibinfo  {journal} {Nature}\ }\textbf {\bibinfo {volume}
  {402}},\ \bibinfo {pages} {390} (\bibinfo {year} {1999})}\BibitemShut
  {NoStop}%
\bibitem [{\citenamefont {Bouwmeester}\ \emph {et~al.}(1997)\citenamefont
  {Bouwmeester}, \citenamefont {Pan}, \citenamefont {Mattle}, \citenamefont
  {Eibl}, \citenamefont {Weinfurter},\ and\ \citenamefont
  {Zeilinger}}]{bouwmeester1997}%
  \BibitemOpen
  \bibfield  {author} {\bibinfo {author} {\bibfnamefont {D.}~\bibnamefont
  {Bouwmeester}}, \bibinfo {author} {\bibfnamefont {J.-W.}\ \bibnamefont
  {Pan}}, \bibinfo {author} {\bibfnamefont {K.}~\bibnamefont {Mattle}},
  \bibinfo {author} {\bibfnamefont {M.}~\bibnamefont {Eibl}}, \bibinfo {author}
  {\bibfnamefont {H.}~\bibnamefont {Weinfurter}},\ and\ \bibinfo {author}
  {\bibfnamefont {A.}~\bibnamefont {Zeilinger}},\ }\bibfield  {title} {\bibinfo
  {title} {Experimental quantum teleportation},\ }\href
  {https://doi.org/10.1038/37539} {\bibfield  {journal} {\bibinfo  {journal}
  {Nature}\ }\textbf {\bibinfo {volume} {390}},\ \bibinfo {pages} {575}
  (\bibinfo {year} {1997})}\BibitemShut {NoStop}%
\bibitem [{\citenamefont {Makarov}\ \emph {et~al.}(2024)\citenamefont
  {Makarov}, \citenamefont {Abrikosov}, \citenamefont {Chaiwongkhot},
  \citenamefont {Fedorov}, \citenamefont {Huang}, \citenamefont {Kiktenko},
  \citenamefont {Petrov}, \citenamefont {Ponosova}, \citenamefont
  {Ruzhitskaya}, \citenamefont {Sajeed}, \citenamefont {Tayduganov},
  \citenamefont {Trefilov},\ and\ \citenamefont {Zaitsev}}]{makarov2024}%
  \BibitemOpen
  \bibfield  {author} {\bibinfo {author} {\bibfnamefont {V.}~\bibnamefont
  {Makarov}}, \bibinfo {author} {\bibfnamefont {A.}~\bibnamefont {Abrikosov}},
  \bibinfo {author} {\bibfnamefont {P.}~\bibnamefont {Chaiwongkhot}}, \bibinfo
  {author} {\bibfnamefont {A.}~\bibnamefont {Fedorov}}, \bibinfo {author}
  {\bibfnamefont {A.}~\bibnamefont {Huang}}, \bibinfo {author} {\bibfnamefont
  {E.}~\bibnamefont {Kiktenko}}, \bibinfo {author} {\bibfnamefont
  {M.}~\bibnamefont {Petrov}}, \bibinfo {author} {\bibfnamefont
  {A.}~\bibnamefont {Ponosova}}, \bibinfo {author} {\bibfnamefont
  {D.}~\bibnamefont {Ruzhitskaya}}, \bibinfo {author} {\bibfnamefont
  {S.}~\bibnamefont {Sajeed}}, \bibinfo {author} {\bibfnamefont
  {A.}~\bibnamefont {Tayduganov}}, \bibinfo {author} {\bibfnamefont
  {D.}~\bibnamefont {Trefilov}},\ and\ \bibinfo {author} {\bibfnamefont
  {K.}~\bibnamefont {Zaitsev}},\ }\bibfield  {title} {\bibinfo {title}
  {Preparing a commercial quantum key distribution system for certification
  against implementation loopholes},\ }\href
  {https://doi.org/10.1103/PhysRevApplied.22.044076} {\bibfield  {journal}
  {\bibinfo  {journal} {Phys. Rev. Appl.}\ }\textbf {\bibinfo {volume} {22}},\
  \bibinfo {pages} {044076} (\bibinfo {year} {2024})}\BibitemShut {NoStop}%
\bibitem [{\citenamefont {Wiechers}\ \emph {et~al.}(2011)\citenamefont
  {Wiechers}, \citenamefont {Lydersen}, \citenamefont {Wittmann}, \citenamefont
  {Elser}, \citenamefont {Skaar}, \citenamefont {Marquardt}, \citenamefont
  {Makarov},\ and\ \citenamefont {Leuchs}}]{wiechers2011}%
  \BibitemOpen
  \bibfield  {author} {\bibinfo {author} {\bibfnamefont {C.}~\bibnamefont
  {Wiechers}}, \bibinfo {author} {\bibfnamefont {L.}~\bibnamefont {Lydersen}},
  \bibinfo {author} {\bibfnamefont {C.}~\bibnamefont {Wittmann}}, \bibinfo
  {author} {\bibfnamefont {D.}~\bibnamefont {Elser}}, \bibinfo {author}
  {\bibfnamefont {J.}~\bibnamefont {Skaar}}, \bibinfo {author} {\bibfnamefont
  {C.}~\bibnamefont {Marquardt}}, \bibinfo {author} {\bibfnamefont
  {V.}~\bibnamefont {Makarov}},\ and\ \bibinfo {author} {\bibfnamefont
  {G.}~\bibnamefont {Leuchs}},\ }\bibfield  {title} {\bibinfo {title}
  {After-gate attack on a quantum cryptosystem},\ }\href
  {https://doi.org/10.1088/1367-2630/13/1/013043} {\bibfield  {journal}
  {\bibinfo  {journal} {New J. Phys.}\ }\textbf {\bibinfo {volume} {13}},\
  \bibinfo {pages} {013043} (\bibinfo {year} {2011})}\BibitemShut {NoStop}%
\bibitem [{\citenamefont {Lydersen}\ \emph
  {et~al.}(2011{\natexlab{a}})\citenamefont {Lydersen}, \citenamefont {Jain},
  \citenamefont {Wittmann}, \citenamefont {Mar{\o}y}, \citenamefont {Skaar},
  \citenamefont {Marquardt}, \citenamefont {Makarov},\ and\ \citenamefont
  {Leuchs}}]{lydersen2011b}%
  \BibitemOpen
  \bibfield  {author} {\bibinfo {author} {\bibfnamefont {L.}~\bibnamefont
  {Lydersen}}, \bibinfo {author} {\bibfnamefont {N.}~\bibnamefont {Jain}},
  \bibinfo {author} {\bibfnamefont {C.}~\bibnamefont {Wittmann}}, \bibinfo
  {author} {\bibfnamefont {{\O}.}~\bibnamefont {Mar{\o}y}}, \bibinfo {author}
  {\bibfnamefont {J.}~\bibnamefont {Skaar}}, \bibinfo {author} {\bibfnamefont
  {C.}~\bibnamefont {Marquardt}}, \bibinfo {author} {\bibfnamefont
  {V.}~\bibnamefont {Makarov}},\ and\ \bibinfo {author} {\bibfnamefont
  {G.}~\bibnamefont {Leuchs}},\ }\bibfield  {title} {\bibinfo {title}
  {Superlinear threshold detectors in quantum cryptography},\ }\href
  {https://doi.org/10.1103/PhysRevA.84.032320} {\bibfield  {journal} {\bibinfo
  {journal} {Phys. Rev. A}\ }\textbf {\bibinfo {volume} {84}},\ \bibinfo
  {pages} {032320} (\bibinfo {year} {2011}{\natexlab{a}})}\BibitemShut
  {NoStop}%
\bibitem [{\citenamefont {Lydersen}\ \emph
  {et~al.}(2011{\natexlab{b}})\citenamefont {Lydersen}, \citenamefont
  {Akhlaghi}, \citenamefont {Majedi}, \citenamefont {Skaar},\ and\
  \citenamefont {Makarov}}]{lydersen2011c}%
  \BibitemOpen
  \bibfield  {author} {\bibinfo {author} {\bibfnamefont {L.}~\bibnamefont
  {Lydersen}}, \bibinfo {author} {\bibfnamefont {M.~K.}\ \bibnamefont
  {Akhlaghi}}, \bibinfo {author} {\bibfnamefont {A.~H.}\ \bibnamefont
  {Majedi}}, \bibinfo {author} {\bibfnamefont {J.}~\bibnamefont {Skaar}},\ and\
  \bibinfo {author} {\bibfnamefont {V.}~\bibnamefont {Makarov}},\ }\bibfield
  {title} {\bibinfo {title} {Controlling a superconducting nanowire
  single-photon detector using tailored bright illumination},\ }\href
  {https://doi.org/10.1088/1367-2630/13/11/113042} {\bibfield  {journal}
  {\bibinfo  {journal} {New J. Phys.}\ }\textbf {\bibinfo {volume} {13}},\
  \bibinfo {pages} {113042} (\bibinfo {year} {2011}{\natexlab{b}})}\BibitemShut
  {NoStop}%
\bibitem [{\citenamefont {Chaiwongkhot}\ \emph {et~al.}(2022)\citenamefont
  {Chaiwongkhot}, \citenamefont {Zhong}, \citenamefont {Huang}, \citenamefont
  {Qin}, \citenamefont {Shi},\ and\ \citenamefont
  {Makarov}}]{chaiwongkhot2022}%
  \BibitemOpen
  \bibfield  {author} {\bibinfo {author} {\bibfnamefont {P.}~\bibnamefont
  {Chaiwongkhot}}, \bibinfo {author} {\bibfnamefont {J.}~\bibnamefont {Zhong}},
  \bibinfo {author} {\bibfnamefont {A.}~\bibnamefont {Huang}}, \bibinfo
  {author} {\bibfnamefont {H.}~\bibnamefont {Qin}}, \bibinfo {author}
  {\bibfnamefont {S.}~\bibnamefont {Shi}},\ and\ \bibinfo {author}
  {\bibfnamefont {V.}~\bibnamefont {Makarov}},\ }\bibfield  {title} {\bibinfo
  {title} {Faking photon number on a transition-edge sensor},\ }\href
  {https://doi.org/10.1140/epjqt/s40507-022-00141-2} {\bibfield  {journal}
  {\bibinfo  {journal} {EPJ Quantum Technol.}\ }\textbf {\bibinfo {volume}
  {9}},\ \bibinfo {pages} {23} (\bibinfo {year} {2022})}\BibitemShut {NoStop}%
\bibitem [{\citenamefont {Tsurumaru}\ and\ \citenamefont
  {Tamaki}(2008)}]{tsurumaru2008}%
  \BibitemOpen
  \bibfield  {author} {\bibinfo {author} {\bibfnamefont {T.}~\bibnamefont
  {Tsurumaru}}\ and\ \bibinfo {author} {\bibfnamefont {K.}~\bibnamefont
  {Tamaki}},\ }\bibfield  {title} {\bibinfo {title} {Security proof for
  quantum-key-distribution systems with threshold detectors},\ }\href
  {https://doi.org/10.1103/PhysRevA.78.032302} {\bibfield  {journal} {\bibinfo
  {journal} {Phys. Rev. A}\ }\textbf {\bibinfo {volume} {78}},\ \bibinfo {eid}
  {032302} (\bibinfo {year} {2008})}\BibitemShut {NoStop}%
\bibitem [{\citenamefont {Xu}\ \emph {et~al.}(2020)\citenamefont {Xu},
  \citenamefont {Ma}, \citenamefont {Zhang}, \citenamefont {Lo},\ and\
  \citenamefont {Pan}}]{xu2020}%
  \BibitemOpen
  \bibfield  {author} {\bibinfo {author} {\bibfnamefont {F.}~\bibnamefont
  {Xu}}, \bibinfo {author} {\bibfnamefont {X.}~\bibnamefont {Ma}}, \bibinfo
  {author} {\bibfnamefont {Q.}~\bibnamefont {Zhang}}, \bibinfo {author}
  {\bibfnamefont {H.-K.}\ \bibnamefont {Lo}},\ and\ \bibinfo {author}
  {\bibfnamefont {J.-W.}\ \bibnamefont {Pan}},\ }\bibfield  {title} {\bibinfo
  {title} {Secure quantum key distribution with realistic devices},\ }\href
  {https://doi.org/10.1103/RevModPhys.92.025002} {\bibfield  {journal}
  {\bibinfo  {journal} {Rev. Mod. Phys.}\ }\textbf {\bibinfo {volume} {92}},\
  \bibinfo {pages} {025002} (\bibinfo {year} {2020})}\BibitemShut {NoStop}%
\bibitem [{iso()}]{iso23837-2023}%
  \BibitemOpen
  \href@noop {} {\bibinfo {title} {{ISO/IEC 23837-2:2023(en).} {I}nformation
  security --- {S}ecurity requirements, test and evaluation methods for quantum
  key distribution --- {P}art~2: {E}valuation and testing methods}},\ \bibinfo
  {note}
  {\url{https://www.iso.org/obp/ui/en/\#iso:std:iso-iec:23837:-2:ed-1:v1:en},
  visited 8 Mar 2025}\BibitemShut {NoStop}%
\bibitem [{\citenamefont {Marquardt}\ \emph {et~al.}()\citenamefont
  {Marquardt}, \citenamefont {Seyfarth}, \citenamefont {Bettendorf},
  \citenamefont {Bohmann}, \citenamefont {Buchner}, \citenamefont {Curty},
  \citenamefont {Elser}, \citenamefont {Eul}, \citenamefont {Gehring},
  \citenamefont {Jain}, \citenamefont {Klocke}, \citenamefont {Reinecke},
  \citenamefont {Sieber}, \citenamefont {Ursin}, \citenamefont {Wehling},\ and\
  \citenamefont {Weier}}]{marquardt2023}%
  \BibitemOpen
  \bibfield  {author} {\bibinfo {author} {\bibfnamefont {C.}~\bibnamefont
  {Marquardt}}, \bibinfo {author} {\bibfnamefont {U.}~\bibnamefont {Seyfarth}},
  \bibinfo {author} {\bibfnamefont {S.}~\bibnamefont {Bettendorf}}, \bibinfo
  {author} {\bibfnamefont {M.}~\bibnamefont {Bohmann}}, \bibinfo {author}
  {\bibfnamefont {A.}~\bibnamefont {Buchner}}, \bibinfo {author} {\bibfnamefont
  {M.}~\bibnamefont {Curty}}, \bibinfo {author} {\bibfnamefont
  {D.}~\bibnamefont {Elser}}, \bibinfo {author} {\bibfnamefont
  {S.}~\bibnamefont {Eul}}, \bibinfo {author} {\bibfnamefont {T.}~\bibnamefont
  {Gehring}}, \bibinfo {author} {\bibfnamefont {N.}~\bibnamefont {Jain}},
  \bibinfo {author} {\bibfnamefont {T.}~\bibnamefont {Klocke}}, \bibinfo
  {author} {\bibfnamefont {M.}~\bibnamefont {Reinecke}}, \bibinfo {author}
  {\bibfnamefont {N.}~\bibnamefont {Sieber}}, \bibinfo {author} {\bibfnamefont
  {R.}~\bibnamefont {Ursin}}, \bibinfo {author} {\bibfnamefont
  {M.}~\bibnamefont {Wehling}},\ and\ \bibinfo {author} {\bibfnamefont
  {H.}~\bibnamefont {Weier}},\ }\href@noop {} {\bibinfo {title} {Implementation
  attacks against {QKD} systems}},\ \bibinfo {note} {{BSI} technical report,
  \url{https://www.bsi.bund.de/EN/Service-Navi/Publikationen/Studien/QKD-Systems/Implementation_Attacks_QKD_Systems_node.html},
  visited 8 Mar 2025}\BibitemShut {NoStop}%
\bibitem [{\citenamefont {Acheva}\ \emph {et~al.}(2023)\citenamefont {Acheva},
  \citenamefont {Zaitsev}, \citenamefont {Zavodilenko}, \citenamefont {Losev},
  \citenamefont {Huang},\ and\ \citenamefont {Makarov}}]{acheva2023}%
  \BibitemOpen
  \bibfield  {author} {\bibinfo {author} {\bibfnamefont {P.}~\bibnamefont
  {Acheva}}, \bibinfo {author} {\bibfnamefont {K.}~\bibnamefont {Zaitsev}},
  \bibinfo {author} {\bibfnamefont {V.}~\bibnamefont {Zavodilenko}}, \bibinfo
  {author} {\bibfnamefont {A.}~\bibnamefont {Losev}}, \bibinfo {author}
  {\bibfnamefont {A.}~\bibnamefont {Huang}},\ and\ \bibinfo {author}
  {\bibfnamefont {V.}~\bibnamefont {Makarov}},\ }\bibfield  {title} {\bibinfo
  {title} {Automated verification of countermeasure against detector-control
  attack in quantum key distribution},\ }\href
  {https://doi.org/10.1140/epjqt/s40507-023-00178-x} {\bibfield  {journal}
  {\bibinfo  {journal} {EPJ Quantum Technol.}\ }\textbf {\bibinfo {volume}
  {10}},\ \bibinfo {pages} {22} (\bibinfo {year} {2023})}\BibitemShut {NoStop}%
\bibitem [{\citenamefont {Losev}\ \emph {et~al.}(2022)\citenamefont {Losev},
  \citenamefont {Zavodilenko}, \citenamefont {Koziy}, \citenamefont
  {Kurochkin},\ and\ \citenamefont {Gorbatsevich}}]{losev2022}%
  \BibitemOpen
  \bibfield  {author} {\bibinfo {author} {\bibfnamefont {A.}~\bibnamefont
  {Losev}}, \bibinfo {author} {\bibfnamefont {V.}~\bibnamefont {Zavodilenko}},
  \bibinfo {author} {\bibfnamefont {A.}~\bibnamefont {Koziy}}, \bibinfo
  {author} {\bibfnamefont {Y.}~\bibnamefont {Kurochkin}},\ and\ \bibinfo
  {author} {\bibfnamefont {A.}~\bibnamefont {Gorbatsevich}},\ }\bibfield
  {title} {\bibinfo {title} {Dependence of functional parameters of sine-gated
  {InGaAs/InP} single-photon avalanche diodes on the gating parameters},\
  }\href {https://doi.org/10.1109/JPHOT.2022.3148204} {\bibfield  {journal}
  {\bibinfo  {journal} {IEEE Photonics J.}\ }\textbf {\bibinfo {volume} {14}},\
  \bibinfo {pages} {6817109} (\bibinfo {year} {2022})}\BibitemShut {NoStop}%
\bibitem [{\citenamefont {Bochkov}\ and\ \citenamefont
  {Trushechkin}(2019)}]{bochkov2019}%
  \BibitemOpen
  \bibfield  {author} {\bibinfo {author} {\bibfnamefont {M.~K.}\ \bibnamefont
  {Bochkov}}\ and\ \bibinfo {author} {\bibfnamefont {A.~S.}\ \bibnamefont
  {Trushechkin}},\ }\bibfield  {title} {\bibinfo {title} {Security of quantum
  key distribution with detection-efficiency mismatch in the single-photon
  case: Tight bounds},\ }\href {https://doi.org/10.1103/PhysRevA.99.032308}
  {\bibfield  {journal} {\bibinfo  {journal} {Phys. Rev. A}\ }\textbf {\bibinfo
  {volume} {99}},\ \bibinfo {pages} {032308} (\bibinfo {year}
  {2019})}\BibitemShut {NoStop}%
\bibitem [{\citenamefont {Trushechkin}(2022)}]{trushechkin2022}%
  \BibitemOpen
  \bibfield  {author} {\bibinfo {author} {\bibfnamefont {A.}~\bibnamefont
  {Trushechkin}},\ }\bibfield  {title} {\bibinfo {title} {Security of quantum
  key distribution with detection-efficiency mismatch in the multiphoton
  case},\ }\href {https://doi.org/10.22331/q-2022-07-22-771} {\bibfield
  {journal} {\bibinfo  {journal} {Quantum}\ }\textbf {\bibinfo {volume} {6}},\
  \bibinfo {pages} {771} (\bibinfo {year} {2022})}\BibitemShut {NoStop}%
\bibitem [{\citenamefont {Koehler-Sidki}\ \emph {et~al.}(2019)\citenamefont
  {Koehler-Sidki}, \citenamefont {Dynes}, \citenamefont {Martinez},
  \citenamefont {Lucamarini}, \citenamefont {Roberts}, \citenamefont {Sharpe},
  \citenamefont {Yuan},\ and\ \citenamefont {Shields}}]{koehler-sidki2019}%
  \BibitemOpen
  \bibfield  {author} {\bibinfo {author} {\bibfnamefont {A.}~\bibnamefont
  {Koehler-Sidki}}, \bibinfo {author} {\bibfnamefont {J.~F.}\ \bibnamefont
  {Dynes}}, \bibinfo {author} {\bibfnamefont {A.}~\bibnamefont {Martinez}},
  \bibinfo {author} {\bibfnamefont {M.}~\bibnamefont {Lucamarini}}, \bibinfo
  {author} {\bibfnamefont {G.~L.}\ \bibnamefont {Roberts}}, \bibinfo {author}
  {\bibfnamefont {A.~W.}\ \bibnamefont {Sharpe}}, \bibinfo {author}
  {\bibfnamefont {Z.~L.}\ \bibnamefont {Yuan}},\ and\ \bibinfo {author}
  {\bibfnamefont {A.}~\bibnamefont {Shields}},\ }\bibfield  {title} {\bibinfo
  {title} {Intrinsic mitigation of the after-gate attack in quantum key
  distribution through fast-gated delayed detection},\ }\href
  {https://doi.org/10.1103/PhysRevApplied.12.024050} {\bibfield  {journal}
  {\bibinfo  {journal} {Phys. Rev. Appl.}\ }\textbf {\bibinfo {volume} {12}},\
  \bibinfo {pages} {024050} (\bibinfo {year} {2019})}\BibitemShut {NoStop}%
\bibitem [{\citenamefont {Ware}\ \emph {et~al.}(2007)\citenamefont {Ware},
  \citenamefont {Migdall}, \citenamefont {Bienfang},\ and\ \citenamefont
  {Polyakov}}]{ware2007}%
  \BibitemOpen
  \bibfield  {author} {\bibinfo {author} {\bibfnamefont {M.}~\bibnamefont
  {Ware}}, \bibinfo {author} {\bibfnamefont {A.}~\bibnamefont {Migdall}},
  \bibinfo {author} {\bibfnamefont {J.~C.}\ \bibnamefont {Bienfang}},\ and\
  \bibinfo {author} {\bibfnamefont {S.~V.}\ \bibnamefont {Polyakov}},\
  }\bibfield  {title} {\bibinfo {title} {Calibrating photon-counting detectors
  to high accuracy: background and deadtime issues},\ }\href
  {https://doi.org/10.1080/09500340600759597} {\bibfield  {journal} {\bibinfo
  {journal} {J. Mod. Opt.}\ }\textbf {\bibinfo {volume} {54}},\ \bibinfo
  {pages} {361} (\bibinfo {year} {2007})}\BibitemShut {NoStop}%
\bibitem [{\citenamefont {Sauge}\ \emph {et~al.}(2011)\citenamefont {Sauge},
  \citenamefont {Lydersen}, \citenamefont {Anisimov}, \citenamefont {Skaar},\
  and\ \citenamefont {Makarov}}]{sauge2011}%
  \BibitemOpen
  \bibfield  {author} {\bibinfo {author} {\bibfnamefont {S.}~\bibnamefont
  {Sauge}}, \bibinfo {author} {\bibfnamefont {L.}~\bibnamefont {Lydersen}},
  \bibinfo {author} {\bibfnamefont {A.}~\bibnamefont {Anisimov}}, \bibinfo
  {author} {\bibfnamefont {J.}~\bibnamefont {Skaar}},\ and\ \bibinfo {author}
  {\bibfnamefont {V.}~\bibnamefont {Makarov}},\ }\bibfield  {title} {\bibinfo
  {title} {Controlling an actively-quenched single photon detector with bright
  light},\ }\href {https://doi.org/10.1364/OE.19.023590} {\bibfield  {journal}
  {\bibinfo  {journal} {Opt. Express}\ }\textbf {\bibinfo {volume} {19}},\
  \bibinfo {pages} {23590} (\bibinfo {year} {2011})}\BibitemShut {NoStop}%
\bibitem [{\citenamefont {Raupach}\ \emph {et~al.}(2022)\citenamefont
  {Raupach}, \citenamefont {Degiovanni}, \citenamefont {Georgieva},
  \citenamefont {Meda}, \citenamefont {Hofer}, \citenamefont {Gramegna},
  \citenamefont {Genovese}, \citenamefont {K{\" u}ck},\ and\ \citenamefont
  {L{\' o}pez}}]{raupach2022}%
  \BibitemOpen
  \bibfield  {author} {\bibinfo {author} {\bibfnamefont {S.~M.~F.}\
  \bibnamefont {Raupach}}, \bibinfo {author} {\bibfnamefont {I.~P.}\
  \bibnamefont {Degiovanni}}, \bibinfo {author} {\bibfnamefont
  {H.}~\bibnamefont {Georgieva}}, \bibinfo {author} {\bibfnamefont
  {A.}~\bibnamefont {Meda}}, \bibinfo {author} {\bibfnamefont {H.}~\bibnamefont
  {Hofer}}, \bibinfo {author} {\bibfnamefont {M.}~\bibnamefont {Gramegna}},
  \bibinfo {author} {\bibfnamefont {M.}~\bibnamefont {Genovese}}, \bibinfo
  {author} {\bibfnamefont {S.}~\bibnamefont {K{\" u}ck}},\ and\ \bibinfo
  {author} {\bibfnamefont {M.}~\bibnamefont {L{\' o}pez}},\ }\bibfield  {title}
  {\bibinfo {title} {Detection rate dependence of the inherent detection
  efficiency in single-photon detectors based on avalanche diodes},\ }\href
  {https://doi.org/10.1103/PhysRevA.105.042615} {\bibfield  {journal} {\bibinfo
   {journal} {Phys. Rev. A}\ }\textbf {\bibinfo {volume} {105}},\ \bibinfo
  {pages} {042615} (\bibinfo {year} {2022})}\BibitemShut {NoStop}%
\bibitem [{\citenamefont {Kurochkin}\ \emph {et~al.}()\citenamefont
  {Kurochkin}, \citenamefont {Papadovasilakis}, \citenamefont {Trushechkin},
  \citenamefont {Piera},\ and\ \citenamefont {Grieve}}]{kurochkin2024}%
  \BibitemOpen
  \bibfield  {author} {\bibinfo {author} {\bibfnamefont {Y.}~\bibnamefont
  {Kurochkin}}, \bibinfo {author} {\bibfnamefont {M.}~\bibnamefont
  {Papadovasilakis}}, \bibinfo {author} {\bibfnamefont {A.}~\bibnamefont
  {Trushechkin}}, \bibinfo {author} {\bibfnamefont {R.}~\bibnamefont {Piera}},\
  and\ \bibinfo {author} {\bibfnamefont {J.~A.}\ \bibnamefont {Grieve}},\
  }\bibfield  {title} {\bibinfo {title} {A practical transmitter device for
  passive state {BB84} quantum key distribution},\ }\href@noop {} {\ }\Eprint
  {https://arxiv.org/abs/2405.08481} {arXiv:2405.08481 [quant-ph]} \BibitemShut
  {NoStop}%
\bibitem [{\citenamefont {L{\" u}tkenhaus}(2002)}]{lutkenhaus2002}%
  \BibitemOpen
  \bibfield  {author} {\bibinfo {author} {\bibfnamefont {N.}~\bibnamefont {L{\"
  u}tkenhaus}},\ }\bibfield  {title} {\bibinfo {title} {Quantum key
  distribution with realistic states: photon-number statistics in the
  photon-number splitting attack},\ }\href
  {https://doi.org/10.1088/1367-2630/4/1/344} {\bibfield  {journal} {\bibinfo
  {journal} {New J. Phys.}\ }\textbf {\bibinfo {volume} {4}},\ \bibinfo {pages}
  {44} (\bibinfo {year} {2002})}\BibitemShut {NoStop}%
\bibitem [{\citenamefont {Stasi}\ \emph {et~al.}(2025)\citenamefont {Stasi},
  \citenamefont {Taher}, \citenamefont {Resta}, \citenamefont {Zbinden},
  \citenamefont {Thew},\ and\ \citenamefont {Bussi\`{e}res}}]{lorenzo2025}%
  \BibitemOpen
  \bibfield  {author} {\bibinfo {author} {\bibfnamefont {L.}~\bibnamefont
  {Stasi}}, \bibinfo {author} {\bibfnamefont {T.}~\bibnamefont {Taher}},
  \bibinfo {author} {\bibfnamefont {G.~V.}\ \bibnamefont {Resta}}, \bibinfo
  {author} {\bibfnamefont {H.}~\bibnamefont {Zbinden}}, \bibinfo {author}
  {\bibfnamefont {R.}~\bibnamefont {Thew}},\ and\ \bibinfo {author}
  {\bibfnamefont {F.}~\bibnamefont {Bussi\`{e}res}},\ }\bibfield  {title}
  {\bibinfo {title} {Enhanced detection rate and high photon-number
  efficiencies with a scalable parallel {SNSPD}},\ }\href
  {https://doi.org/10.1021/acsphotonics.4c01680} {\bibfield  {journal}
  {\bibinfo  {journal} {ASC Photonics}\ }\textbf {\bibinfo {volume} {12}},\
  \bibinfo {pages} {320} (\bibinfo {year} {2025})}\BibitemShut {NoStop}%
\bibitem [{idq()}]{idqsnspd}%
  \BibitemOpen
  \href@noop {} {\bibinfo {title} {{ID Q}uantique, {P}hoton-number-resolving
  detection}},\ \bibinfo {note}
  {\url{https://www.idquantique.com/quantum-detection-systems/photon-number-resolving-detection/},
  visited 8 Mar 2025}\BibitemShut {NoStop}%
\bibitem [{\citenamefont {Fei}\ \emph {et~al.}(2018)\citenamefont {Fei},
  \citenamefont {Meng}, \citenamefont {Gao}, \citenamefont {Wang},\ and\
  \citenamefont {Ma}}]{fei2018}%
  \BibitemOpen
  \bibfield  {author} {\bibinfo {author} {\bibfnamefont {Y.-Y.}\ \bibnamefont
  {Fei}}, \bibinfo {author} {\bibfnamefont {X.-D.}\ \bibnamefont {Meng}},
  \bibinfo {author} {\bibfnamefont {M.}~\bibnamefont {Gao}}, \bibinfo {author}
  {\bibfnamefont {H.}~\bibnamefont {Wang}},\ and\ \bibinfo {author}
  {\bibfnamefont {Z.}~\bibnamefont {Ma}},\ }\bibfield  {title} {\bibinfo
  {title} {Quantum man-in-the-middle attack on the calibration process of
  quantum key distribution},\ }\href
  {https://doi.org/10.1038/s41598-018-22700-3} {\bibfield  {journal} {\bibinfo
  {journal} {Sci. Rep.}\ }\textbf {\bibinfo {volume} {8}},\ \bibinfo {pages}
  {4283} (\bibinfo {year} {2018})}\BibitemShut {NoStop}%
\bibitem [{\citenamefont {Gottesman}\ \emph {et~al.}(2004)\citenamefont
  {Gottesman}, \citenamefont {Lo}, \citenamefont {L{\" u}tkenhaus},\ and\
  \citenamefont {Preskill}}]{gottesman2004}%
  \BibitemOpen
  \bibfield  {author} {\bibinfo {author} {\bibfnamefont {D.}~\bibnamefont
  {Gottesman}}, \bibinfo {author} {\bibfnamefont {H.-K.}\ \bibnamefont {Lo}},
  \bibinfo {author} {\bibfnamefont {N.}~\bibnamefont {L{\" u}tkenhaus}},\ and\
  \bibinfo {author} {\bibfnamefont {J.}~\bibnamefont {Preskill}},\ }\bibfield
  {title} {\bibinfo {title} {Security of quantum key distribution with
  imperfect devices},\ }\href@noop {} {\bibfield  {journal} {\bibinfo
  {journal} {Quantum Inf. Comput.}\ }\textbf {\bibinfo {volume} {4}},\ \bibinfo
  {pages} {325} (\bibinfo {year} {2004})}\BibitemShut {NoStop}%
\bibitem [{\citenamefont {Hwang}(2003)}]{hwang2003}%
  \BibitemOpen
  \bibfield  {author} {\bibinfo {author} {\bibfnamefont {W.-Y.}\ \bibnamefont
  {Hwang}},\ }\bibfield  {title} {\bibinfo {title} {Quantum key distribution
  with high loss: Toward global secure communication},\ }\href
  {https://doi.org/10.1103/PhysRevLett.91.057901} {\bibfield  {journal}
  {\bibinfo  {journal} {Phys. Rev. Lett.}\ }\textbf {\bibinfo {volume} {91}},\
  \bibinfo {pages} {057901} (\bibinfo {year} {2003})}\BibitemShut {NoStop}%
\bibitem [{\citenamefont {Scarani}\ \emph {et~al.}(2009)\citenamefont
  {Scarani}, \citenamefont {Bechmann-Pasquinucci}, \citenamefont {Cerf},
  \citenamefont {Du\v{s}ek}, \citenamefont {L\"{u}tkenhaus},\ and\
  \citenamefont {Peev}}]{scarani2009}%
  \BibitemOpen
  \bibfield  {author} {\bibinfo {author} {\bibfnamefont {V.}~\bibnamefont
  {Scarani}}, \bibinfo {author} {\bibfnamefont {H.}~\bibnamefont
  {Bechmann-Pasquinucci}}, \bibinfo {author} {\bibfnamefont {N.~J.}\
  \bibnamefont {Cerf}}, \bibinfo {author} {\bibfnamefont {M.}~\bibnamefont
  {Du\v{s}ek}}, \bibinfo {author} {\bibfnamefont {N.}~\bibnamefont
  {L\"{u}tkenhaus}},\ and\ \bibinfo {author} {\bibfnamefont {M.}~\bibnamefont
  {Peev}},\ }\bibfield  {title} {\bibinfo {title} {The security of practical
  quantum key distribution},\ }\href
  {https://doi.org/10.1103/RevModPhys.81.1301} {\bibfield  {journal} {\bibinfo
  {journal} {Rev. Mod. Phys.}\ }\textbf {\bibinfo {volume} {81}},\ \bibinfo
  {eid} {1301} (\bibinfo {year} {2009})}\BibitemShut {NoStop}%
\bibitem [{\citenamefont {F{\' e}lix}\ \emph {et~al.}(2001)\citenamefont {F{\'
  e}lix}, \citenamefont {Gisin}, \citenamefont {Stefanov},\ and\ \citenamefont
  {Zbinden}}]{felix2001}%
  \BibitemOpen
  \bibfield  {author} {\bibinfo {author} {\bibfnamefont {S.}~\bibnamefont {F{\'
  e}lix}}, \bibinfo {author} {\bibfnamefont {N.}~\bibnamefont {Gisin}},
  \bibinfo {author} {\bibfnamefont {A.}~\bibnamefont {Stefanov}},\ and\
  \bibinfo {author} {\bibfnamefont {H.}~\bibnamefont {Zbinden}},\ }\bibfield
  {title} {\bibinfo {title} {Faint laser quantum key distribution:
  eavesdropping exploiting multiphoton pulses},\ }\href
  {https://doi.org/10.1080/09500340108240903} {\bibfield  {journal} {\bibinfo
  {journal} {J. Mod. Opt.}\ }\textbf {\bibinfo {volume} {48}},\ \bibinfo
  {pages} {2009} (\bibinfo {year} {2001})}\BibitemShut {NoStop}%
\bibitem [{\citenamefont {Huang}\ \emph {et~al.}(2026)\citenamefont {Huang},
  \citenamefont {Peng}, \citenamefont {Liu}, \citenamefont {Yuan},
  \citenamefont {Peng}, \citenamefont {Yang}, \citenamefont {Wu}, \citenamefont
  {Chen}, \citenamefont {Sun}, \citenamefont {Wang},\ and\ \citenamefont
  {Makarov}}]{huang2026}%
  \BibitemOpen
  \bibfield  {author} {\bibinfo {author} {\bibfnamefont {A.}~\bibnamefont
  {Huang}}, \bibinfo {author} {\bibfnamefont {Q.}~\bibnamefont {Peng}},
  \bibinfo {author} {\bibfnamefont {J.}~\bibnamefont {Liu}}, \bibinfo {author}
  {\bibfnamefont {X.}~\bibnamefont {Yuan}}, \bibinfo {author} {\bibfnamefont
  {C.}~\bibnamefont {Peng}}, \bibinfo {author} {\bibfnamefont {Z.}~\bibnamefont
  {Yang}}, \bibinfo {author} {\bibfnamefont {H.}~\bibnamefont {Wu}}, \bibinfo
  {author} {\bibfnamefont {Z.}~\bibnamefont {Chen}}, \bibinfo {author}
  {\bibfnamefont {G.}~\bibnamefont {Sun}}, \bibinfo {author} {\bibfnamefont
  {F.}~\bibnamefont {Wang}},\ and\ \bibinfo {author} {\bibfnamefont
  {V.}~\bibnamefont {Makarov}},\ }\bibfield  {title} {\bibinfo {title}
  {Penetration testing of quantum key distribution system as a black box},\
  }\href {https://doi.org/10.1093/nsr/nwag174} {\bibfield  {journal} {\bibinfo
  {journal} {Natl. Sci. Rev.}\ }\textbf {\bibinfo {volume} {13}},\ \bibinfo
  {pages} {nwag174} (\bibinfo {year} {2026})}\BibitemShut {NoStop}%
\bibitem [{\citenamefont {Wang}\ \emph {et~al.}(2023)\citenamefont {Wang},
  \citenamefont {Zhang}, \citenamefont {Xie},\ and\ \citenamefont
  {Guo}}]{wang2023}%
  \BibitemOpen
  \bibfield  {author} {\bibinfo {author} {\bibfnamefont {P.}~\bibnamefont
  {Wang}}, \bibinfo {author} {\bibfnamefont {Q.}~\bibnamefont {Zhang}},
  \bibinfo {author} {\bibfnamefont {H.}~\bibnamefont {Xie}},\ and\ \bibinfo
  {author} {\bibfnamefont {B.}~\bibnamefont {Guo}},\ }\bibfield  {title}
  {\bibinfo {title} {Optimized polarization encoder with high extinction ratio
  for quantum key distribution system},\ }\href
  {https://doi.org/10.3390/electronics12081859} {\bibfield  {journal} {\bibinfo
   {journal} {Electronics}\ }\textbf {\bibinfo {volume} {12}},\ \bibinfo
  {pages} {1859} (\bibinfo {year} {2023})}\BibitemShut {NoStop}%
\bibitem [{\citenamefont {Bourgoin}\ \emph {et~al.}(2013)\citenamefont
  {Bourgoin}, \citenamefont {Meyer-Scott}, \citenamefont {Higgins},
  \citenamefont {Helou}, \citenamefont {Erven}, \citenamefont {H{\" u}bel},
  \citenamefont {Kumar}, \citenamefont {Hudson}, \citenamefont {D'Souza},
  \citenamefont {Girard}, \citenamefont {Laflamme},\ and\ \citenamefont
  {Jennewein}}]{bourgoin2013}%
  \BibitemOpen
  \bibfield  {author} {\bibinfo {author} {\bibfnamefont {J.-P.}\ \bibnamefont
  {Bourgoin}}, \bibinfo {author} {\bibfnamefont {E.}~\bibnamefont
  {Meyer-Scott}}, \bibinfo {author} {\bibfnamefont {B.~L.}\ \bibnamefont
  {Higgins}}, \bibinfo {author} {\bibfnamefont {B.}~\bibnamefont {Helou}},
  \bibinfo {author} {\bibfnamefont {C.}~\bibnamefont {Erven}}, \bibinfo
  {author} {\bibfnamefont {H.}~\bibnamefont {H{\" u}bel}}, \bibinfo {author}
  {\bibfnamefont {B.}~\bibnamefont {Kumar}}, \bibinfo {author} {\bibfnamefont
  {D.}~\bibnamefont {Hudson}}, \bibinfo {author} {\bibfnamefont
  {I.}~\bibnamefont {D'Souza}}, \bibinfo {author} {\bibfnamefont
  {R.}~\bibnamefont {Girard}}, \bibinfo {author} {\bibfnamefont
  {R.}~\bibnamefont {Laflamme}},\ and\ \bibinfo {author} {\bibfnamefont
  {T.}~\bibnamefont {Jennewein}},\ }\bibfield  {title} {\bibinfo {title} {A
  comprehensive design and performance analysis of low {E}arth orbit satellite
  quantum communication},\ }\href
  {https://doi.org/10.1088/1367-2630/15/2/023006} {\bibfield  {journal}
  {\bibinfo  {journal} {New J. Phys.}\ }\textbf {\bibinfo {volume} {15}},\
  \bibinfo {pages} {023006} (\bibinfo {year} {2013})}\BibitemShut {NoStop}%
\bibitem [{\citenamefont {Liao}\ \emph {et~al.}(2017)\citenamefont {Liao},
  \citenamefont {Cai}, \citenamefont {Liu}, \citenamefont {Zhang},
  \citenamefont {Li}, \citenamefont {Ren}, \citenamefont {Yin}, \citenamefont
  {Shen}, \citenamefont {Cao}, \citenamefont {Li}, \citenamefont {Li},
  \citenamefont {Chen}, \citenamefont {Sun}, \citenamefont {Jia}, \citenamefont
  {Wu}, \citenamefont {Jiang}, \citenamefont {Wang}, \citenamefont {Huang},
  \citenamefont {Wang}, \citenamefont {Zhou}, \citenamefont {Deng},
  \citenamefont {Xi}, \citenamefont {Ma}, \citenamefont {Hu}, \citenamefont
  {Zhang}, \citenamefont {Chen}, \citenamefont {Liu}, \citenamefont {Wang},
  \citenamefont {Zhu}, \citenamefont {Lu}, \citenamefont {Shu}, \citenamefont
  {Peng}, \citenamefont {Wang},\ and\ \citenamefont {Pan}}]{liao2017}%
  \BibitemOpen
  \bibfield  {author} {\bibinfo {author} {\bibfnamefont {S.-K.}\ \bibnamefont
  {Liao}}, \bibinfo {author} {\bibfnamefont {W.-Q.}\ \bibnamefont {Cai}},
  \bibinfo {author} {\bibfnamefont {W.-Y.}\ \bibnamefont {Liu}}, \bibinfo
  {author} {\bibfnamefont {L.}~\bibnamefont {Zhang}}, \bibinfo {author}
  {\bibfnamefont {Y.}~\bibnamefont {Li}}, \bibinfo {author} {\bibfnamefont
  {J.-G.}\ \bibnamefont {Ren}}, \bibinfo {author} {\bibfnamefont
  {J.}~\bibnamefont {Yin}}, \bibinfo {author} {\bibfnamefont {Q.}~\bibnamefont
  {Shen}}, \bibinfo {author} {\bibfnamefont {Y.}~\bibnamefont {Cao}}, \bibinfo
  {author} {\bibfnamefont {Z.-P.}\ \bibnamefont {Li}}, \bibinfo {author}
  {\bibfnamefont {F.-Z.}\ \bibnamefont {Li}}, \bibinfo {author} {\bibfnamefont
  {X.-W.}\ \bibnamefont {Chen}}, \bibinfo {author} {\bibfnamefont {L.-H.}\
  \bibnamefont {Sun}}, \bibinfo {author} {\bibfnamefont {J.-J.}\ \bibnamefont
  {Jia}}, \bibinfo {author} {\bibfnamefont {J.-C.}\ \bibnamefont {Wu}},
  \bibinfo {author} {\bibfnamefont {X.-J.}\ \bibnamefont {Jiang}}, \bibinfo
  {author} {\bibfnamefont {J.-F.}\ \bibnamefont {Wang}}, \bibinfo {author}
  {\bibfnamefont {Y.-M.}\ \bibnamefont {Huang}}, \bibinfo {author}
  {\bibfnamefont {Q.}~\bibnamefont {Wang}}, \bibinfo {author} {\bibfnamefont
  {Y.-L.}\ \bibnamefont {Zhou}}, \bibinfo {author} {\bibfnamefont
  {L.}~\bibnamefont {Deng}}, \bibinfo {author} {\bibfnamefont {T.}~\bibnamefont
  {Xi}}, \bibinfo {author} {\bibfnamefont {L.}~\bibnamefont {Ma}}, \bibinfo
  {author} {\bibfnamefont {T.}~\bibnamefont {Hu}}, \bibinfo {author}
  {\bibfnamefont {Q.}~\bibnamefont {Zhang}}, \bibinfo {author} {\bibfnamefont
  {Y.-A.}\ \bibnamefont {Chen}}, \bibinfo {author} {\bibfnamefont {N.-L.}\
  \bibnamefont {Liu}}, \bibinfo {author} {\bibfnamefont {X.-B.}\ \bibnamefont
  {Wang}}, \bibinfo {author} {\bibfnamefont {Z.-C.}\ \bibnamefont {Zhu}},
  \bibinfo {author} {\bibfnamefont {C.-Y.}\ \bibnamefont {Lu}}, \bibinfo
  {author} {\bibfnamefont {R.}~\bibnamefont {Shu}}, \bibinfo {author}
  {\bibfnamefont {C.-Z.}\ \bibnamefont {Peng}}, \bibinfo {author}
  {\bibfnamefont {J.-Y.}\ \bibnamefont {Wang}},\ and\ \bibinfo {author}
  {\bibfnamefont {J.-W.}\ \bibnamefont {Pan}},\ }\bibfield  {title} {\bibinfo
  {title} {Satellite-to-ground quantum key distribution},\ }\href
  {https://doi.org/10.1038/nature23655} {\bibfield  {journal} {\bibinfo
  {journal} {Nature}\ }\textbf {\bibinfo {volume} {549}},\ \bibinfo {pages}
  {43} (\bibinfo {year} {2017})}\BibitemShut {NoStop}%
\bibitem [{\citenamefont {Li}\ \emph {et~al.}(2025)\citenamefont {Li},
  \citenamefont {Cai}, \citenamefont {Ren}, \citenamefont {Wang}, \citenamefont
  {Yang}, \citenamefont {Zhang}, \citenamefont {Wu}, \citenamefont {Chang},
  \citenamefont {Wu}, \citenamefont {Jin}, \citenamefont {Xue}, \citenamefont
  {Li}, \citenamefont {Liu}, \citenamefont {Yu}, \citenamefont {Tao},
  \citenamefont {Chen}, \citenamefont {Liu}, \citenamefont {Luo}, \citenamefont
  {Zhou}, \citenamefont {Yong}, \citenamefont {Li}, \citenamefont {Li},
  \citenamefont {Jiang}, \citenamefont {Chen}, \citenamefont {Wu},
  \citenamefont {Tong}, \citenamefont {Xie}, \citenamefont {Zhou},
  \citenamefont {Liu}, \citenamefont {Ismail}, \citenamefont {Petruccione},
  \citenamefont {Liu}, \citenamefont {Li}, \citenamefont {Xu}, \citenamefont
  {Cao}, \citenamefont {Yin}, \citenamefont {Shu}, \citenamefont {Wang},
  \citenamefont {Zhang}, \citenamefont {Wang}, \citenamefont {Liao},
  \citenamefont {Peng},\ and\ \citenamefont {Pan}}]{li2025}%
  \BibitemOpen
  \bibfield  {author} {\bibinfo {author} {\bibfnamefont {Y.}~\bibnamefont
  {Li}}, \bibinfo {author} {\bibfnamefont {W.-Q.}\ \bibnamefont {Cai}},
  \bibinfo {author} {\bibfnamefont {J.-G.}\ \bibnamefont {Ren}}, \bibinfo
  {author} {\bibfnamefont {C.-Z.}\ \bibnamefont {Wang}}, \bibinfo {author}
  {\bibfnamefont {M.}~\bibnamefont {Yang}}, \bibinfo {author} {\bibfnamefont
  {L.}~\bibnamefont {Zhang}}, \bibinfo {author} {\bibfnamefont {H.-Y.}\
  \bibnamefont {Wu}}, \bibinfo {author} {\bibfnamefont {L.}~\bibnamefont
  {Chang}}, \bibinfo {author} {\bibfnamefont {J.-C.}\ \bibnamefont {Wu}},
  \bibinfo {author} {\bibfnamefont {B.}~\bibnamefont {Jin}}, \bibinfo {author}
  {\bibfnamefont {H.-J.}\ \bibnamefont {Xue}}, \bibinfo {author} {\bibfnamefont
  {X.-J.}\ \bibnamefont {Li}}, \bibinfo {author} {\bibfnamefont
  {H.}~\bibnamefont {Liu}}, \bibinfo {author} {\bibfnamefont {G.-W.}\
  \bibnamefont {Yu}}, \bibinfo {author} {\bibfnamefont {X.-Y.}\ \bibnamefont
  {Tao}}, \bibinfo {author} {\bibfnamefont {T.}~\bibnamefont {Chen}}, \bibinfo
  {author} {\bibfnamefont {C.-F.}\ \bibnamefont {Liu}}, \bibinfo {author}
  {\bibfnamefont {W.-B.}\ \bibnamefont {Luo}}, \bibinfo {author} {\bibfnamefont
  {J.}~\bibnamefont {Zhou}}, \bibinfo {author} {\bibfnamefont {H.-L.}\
  \bibnamefont {Yong}}, \bibinfo {author} {\bibfnamefont {Y.-H.}\ \bibnamefont
  {Li}}, \bibinfo {author} {\bibfnamefont {F.-Z.}\ \bibnamefont {Li}}, \bibinfo
  {author} {\bibfnamefont {C.}~\bibnamefont {Jiang}}, \bibinfo {author}
  {\bibfnamefont {H.-Z.}\ \bibnamefont {Chen}}, \bibinfo {author}
  {\bibfnamefont {C.}~\bibnamefont {Wu}}, \bibinfo {author} {\bibfnamefont
  {X.-H.}\ \bibnamefont {Tong}}, \bibinfo {author} {\bibfnamefont {S.-J.}\
  \bibnamefont {Xie}}, \bibinfo {author} {\bibfnamefont {F.}~\bibnamefont
  {Zhou}}, \bibinfo {author} {\bibfnamefont {W.-Y.}\ \bibnamefont {Liu}},
  \bibinfo {author} {\bibfnamefont {Y.}~\bibnamefont {Ismail}}, \bibinfo
  {author} {\bibfnamefont {F.}~\bibnamefont {Petruccione}}, \bibinfo {author}
  {\bibfnamefont {N.-L.}\ \bibnamefont {Liu}}, \bibinfo {author} {\bibfnamefont
  {L.}~\bibnamefont {Li}}, \bibinfo {author} {\bibfnamefont {F.}~\bibnamefont
  {Xu}}, \bibinfo {author} {\bibfnamefont {Y.}~\bibnamefont {Cao}}, \bibinfo
  {author} {\bibfnamefont {J.}~\bibnamefont {Yin}}, \bibinfo {author}
  {\bibfnamefont {R.}~\bibnamefont {Shu}}, \bibinfo {author} {\bibfnamefont
  {X.-B.}\ \bibnamefont {Wang}}, \bibinfo {author} {\bibfnamefont
  {Q.}~\bibnamefont {Zhang}}, \bibinfo {author} {\bibfnamefont {J.-Y.}\
  \bibnamefont {Wang}}, \bibinfo {author} {\bibfnamefont {S.-K.}\ \bibnamefont
  {Liao}}, \bibinfo {author} {\bibfnamefont {C.-Z.}\ \bibnamefont {Peng}},\
  and\ \bibinfo {author} {\bibfnamefont {J.-W.}\ \bibnamefont {Pan}},\
  }\bibfield  {title} {\bibinfo {title} {Microsatellite-based real-time quantum
  key distribution},\ }\href {https://doi.org/10.1038/s41586-025-08739-z}
  {\bibfield  {journal} {\bibinfo  {journal} {Nature}\ }\textbf {\bibinfo
  {volume} {640}},\ \bibinfo {pages} {47} (\bibinfo {year} {2025})}\BibitemShut
  {NoStop}%
\bibitem [{\citenamefont {L\"utkenhaus}(1999)}]{lutkenhaus1999}%
  \BibitemOpen
  \bibfield  {author} {\bibinfo {author} {\bibfnamefont {N.}~\bibnamefont
  {L\"utkenhaus}},\ }\bibfield  {title} {\bibinfo {title} {Estimates for
  practical quantum cryptography},\ }\href
  {https://doi.org/10.1103/PhysRevA.59.3301} {\bibfield  {journal} {\bibinfo
  {journal} {Phys. Rev. A}\ }\textbf {\bibinfo {volume} {59}},\ \bibinfo
  {pages} {3301} (\bibinfo {year} {1999})}\BibitemShut {NoStop}%
\bibitem [{\citenamefont {Lydersen}\ \emph
  {et~al.}(2010{\natexlab{b}})\citenamefont {Lydersen}, \citenamefont
  {Wiechers}, \citenamefont {Wittmann}, \citenamefont {Elser}, \citenamefont
  {Skaar},\ and\ \citenamefont {Makarov}}]{lydersen2010b}%
  \BibitemOpen
  \bibfield  {author} {\bibinfo {author} {\bibfnamefont {L.}~\bibnamefont
  {Lydersen}}, \bibinfo {author} {\bibfnamefont {C.}~\bibnamefont {Wiechers}},
  \bibinfo {author} {\bibfnamefont {C.}~\bibnamefont {Wittmann}}, \bibinfo
  {author} {\bibfnamefont {D.}~\bibnamefont {Elser}}, \bibinfo {author}
  {\bibfnamefont {J.}~\bibnamefont {Skaar}},\ and\ \bibinfo {author}
  {\bibfnamefont {V.}~\bibnamefont {Makarov}},\ }\bibfield  {title} {\bibinfo
  {title} {Thermal blinding of gated detectors in quantum cryptography},\
  }\href {https://doi.org/10.1364/oe.18.027938} {\bibfield  {journal} {\bibinfo
   {journal} {Opt. Express}\ }\textbf {\bibinfo {volume} {18}},\ \bibinfo
  {pages} {27938} (\bibinfo {year} {2010}{\natexlab{b}})}\BibitemShut {NoStop}%
\bibitem [{\citenamefont {Lucamarini}\ \emph {et~al.}(2018)\citenamefont
  {Lucamarini}, \citenamefont {Yuan}, \citenamefont {Dynes},\ and\
  \citenamefont {Shields}}]{lucamarini2018}%
  \BibitemOpen
  \bibfield  {author} {\bibinfo {author} {\bibfnamefont {M.}~\bibnamefont
  {Lucamarini}}, \bibinfo {author} {\bibfnamefont {Z.~L.}\ \bibnamefont
  {Yuan}}, \bibinfo {author} {\bibfnamefont {J.~F.}\ \bibnamefont {Dynes}},\
  and\ \bibinfo {author} {\bibfnamefont {A.~J.}\ \bibnamefont {Shields}},\
  }\bibfield  {title} {\bibinfo {title} {Overcoming the rate--distance limit of
  quantum key distribution without quantum repeaters},\ }\href
  {https://doi.org/10.1038/s41586-018-0066-6} {\bibfield  {journal} {\bibinfo
  {journal} {Nature}\ }\textbf {\bibinfo {volume} {557}},\ \bibinfo {pages}
  {400} (\bibinfo {year} {2018})}\BibitemShut {NoStop}%
\bibitem [{\citenamefont {Wang}\ \emph {et~al.}(2018)\citenamefont {Wang},
  \citenamefont {Yu},\ and\ \citenamefont {Hu}}]{wang2018}%
  \BibitemOpen
  \bibfield  {author} {\bibinfo {author} {\bibfnamefont {X.-B.}\ \bibnamefont
  {Wang}}, \bibinfo {author} {\bibfnamefont {Z.-W.}\ \bibnamefont {Yu}},\ and\
  \bibinfo {author} {\bibfnamefont {X.-L.}\ \bibnamefont {Hu}},\ }\bibfield
  {title} {\bibinfo {title} {Twin-field quantum key distribution with large
  misalignment error},\ }\href {https://doi.org/10.1103/PhysRevA.98.062323}
  {\bibfield  {journal} {\bibinfo  {journal} {Phys. Rev. A}\ }\textbf {\bibinfo
  {volume} {98}},\ \bibinfo {pages} {062323} (\bibinfo {year}
  {2018})}\BibitemShut {NoStop}%
\bibitem [{\citenamefont {Gol'tsman}\ \emph {et~al.}(2001)\citenamefont
  {Gol'tsman}, \citenamefont {Okunev}, \citenamefont {Chulkova}, \citenamefont
  {Lipatov}, \citenamefont {Semenov}, \citenamefont {Smirnov}, \citenamefont
  {Voronov}, \citenamefont {Dzardanov}, \citenamefont {Williams},\ and\
  \citenamefont {Sobolewski}}]{gol'tsman2001}%
  \BibitemOpen
  \bibfield  {author} {\bibinfo {author} {\bibfnamefont {G.~N.}\ \bibnamefont
  {Gol'tsman}}, \bibinfo {author} {\bibfnamefont {O.}~\bibnamefont {Okunev}},
  \bibinfo {author} {\bibfnamefont {G.}~\bibnamefont {Chulkova}}, \bibinfo
  {author} {\bibfnamefont {A.}~\bibnamefont {Lipatov}}, \bibinfo {author}
  {\bibfnamefont {A.}~\bibnamefont {Semenov}}, \bibinfo {author} {\bibfnamefont
  {K.}~\bibnamefont {Smirnov}}, \bibinfo {author} {\bibfnamefont
  {B.}~\bibnamefont {Voronov}}, \bibinfo {author} {\bibfnamefont
  {A.}~\bibnamefont {Dzardanov}}, \bibinfo {author} {\bibfnamefont
  {C.}~\bibnamefont {Williams}},\ and\ \bibinfo {author} {\bibfnamefont
  {R.}~\bibnamefont {Sobolewski}},\ }\bibfield  {title} {\bibinfo {title}
  {Picosecond superconducting single-photon optical detector},\ }\href
  {https://doi.org/10.1063/1.1388868} {\bibfield  {journal} {\bibinfo
  {journal} {Appl. Phys. Lett.}\ }\textbf {\bibinfo {volume} {79}},\ \bibinfo
  {pages} {705} (\bibinfo {year} {2001})}\BibitemShut {NoStop}%
\end{thebibliography}
%

\end{document}